\documentclass[prb,superscriptaddress,showpacs,twocolumn]{revtex4}

\usepackage{graphicx}
\usepackage[latin1]{inputenc}
\usepackage{amsmath}

\newcommand{\eq}[1]{Eq.\,(\ref{#1})}

\newcommand{\SKIP}[1]{}

\newcommand{\fig}[1]{Fig.\,\ref{#1}}
\newcommand{\sect}[1]{Sec.\,\ref{#1}}
\newcommand{\R}{{\mathbf R}}
\newcommand{\intra}{{(\mbox{\small c})} }
\newcommand{\inter}{{(\mbox{\small i})}}
\newcommand{\tr}{{\rm tr}\,}

\newcommand{\tabl}[1]{Tab.\,\ref{#1}}
\newcommand{\onlref}[1]{Ref.\,\onlinecite{#1}}

\begin{document}

\title{Charge ordering in extended Hubbard models:
Variational cluster approach}

\author{M. Aichhorn}
\author{H. G. Evertz}
\author{W. von der Linden}
\affiliation{Institut f\"ur Theoretische Physik,
Technische Universit\"at Graz,
Petersgasse 16, A-8010 Graz, Austria}
\author{M. Potthoff}
\affiliation{Institut f\"ur Theoretische Physik und Astrophysik,
  Universit\"at W\"urzburg, Am Hubland, D-97074 W\"urzburg, Germany}


\begin{abstract}

We present a generalization of the recently proposed variational cluster
perturbation theory to extended Hubbard models at half filling with repulsive
nearest neighbor interaction. The method
takes into account short-range correlations correctly by the exact
diagonalisation of clusters of finite size, whereas
long-range order beyond the size of the clusters is treated on a mean-field
level. For one dimension, we show that quantum Monte Carlo and density-matrix 
renormalization-group results can be reproduced with very good
accuracy. Moreover we apply the method to the two-dimensional extended Hubbard model
on a square lattice. In contrast to the one-dimensional case, a first order phase
transition between spin density wave phase and charge density wave phase is
found as function of the nearest-neighbor interaction 
at onsite interactions $U\ge 3t$. The single-particle spectral function is 
calculated for both the one-dimensional and the two-dimensional system.    
 
\end{abstract}


\maketitle

\section{Introduction}

In recent years an increasing number of theoretical and experimental 
studies in condensed matter physics have focused on the description and 
understanding of
quasi one and two dimensional strongly correlated electronic
systems.
Several fascinating properties of these materials are due to the competition
between different phases with long-range order.
High-temperature superconductivity in cuprates is one of the most famous examples 
which is not yet understood in a satisfactory way. Realistic models that are used 
in this context consist of a kinetic part which accounts for the electron motion
and an interaction part which is of the same order of magnitude. 
The simplest model that can be constructed under these assumptions is the 
tight binding Hubbard model. It consists of a
kinetic energy part, where the electrons can only hop between nearest
neighbor sites and the Coulomb interaction $U$ which acts only locally on each
site. Although this model was used with great success for the description of
a wide class of materials, there are interesting physical questions which
require an extension. 
The inclusion of the nearest-neighbor Coulomb interaction, for example, is 
necessary for the study of inhomogeneous phases, such as the charge-density
wave (CDW).
This leads to the so called extended Hubbard model (EHM).

But knowing the appropriate model
for the description of a material is only the first step on the way to
understanding the physics.
Already for the simple Hubbard model without
non-local Coulomb interaction, an exact calculation of static and
dynamic properties is possible in very special cases only and one has
to be content with approximate methods in general.
For the interesting case where the Coulomb interaction $U$ is of the same 
order of magnitude as the bandwith $W$, the conventional perturbative approach
must fail. This is expected for weak-coupling perturbation theory but also
for the complementary approach with exact treatment of the interaction part 
and perturbative treatment of the kinetic energy. \cite{Metzner91,Pairault98,Pairault00}
\SKIP{\cite{Hubbard63}}

Numerical methods are more promising, such as quantum Monte Carlo 
(QMC),\cite{Dagotto94} exact diagonalisation (ED), and
density-matrix renormalization group (DMRG).\cite{White92} 
They are able to give essentially exact results -- at least for 
limited system sizes or (DMRG) for the one dimensional case.
Another non-perturbative approach is the mean-field method and, in the 
context of the Hubbard model, the dynamical mean field theory (DMFT), 
\cite{Georges96} in particular.
While the DMFT directly works in the thermodynamic limit of infinite system 
size, it must be regarded as a strong approximation since spatial correlations 
are neglected altogether.
Cluster generalizations of the DMFT include at least
short-range correlations via the exact treatment of a small cluster 
instead of considering a single impurity only.
Both, a reciprocal-space (dynamical cluster approximation, DCA\cite{Hettler98}) 
and a real-space construction (cellular dynamical mean field theory, C-DMFT 
\cite{Kotliar01,Biroli02,Bolech03}) have been suggested.

Essentially the same idea is followed with the cluster perturbation 
theory (CPT),\cite{Gros93,Senechal00,Senechal02} 
which is a cluster extension of the strong-coupling expansion for the 
Hubbard model: The lattice is divided into small clusters which are solved
exactly while the hopping between adjacent clusters is
treated perturbatively. The lowest order of the strong-coupling expansion 
in the inter-cluster hopping yields the CPT. Short range correlations
on the scale of the cluster are taken into account exactly, for instance by
the Lanczos technique at zero temperature, while correlations on a scale
larger than the cluster size are neglected.
The CPT is a systematic approach with respect to the cluster size, i.e.\ the 
method becomes exact in the limit $N_c\to\infty$, where $N_c$ is the
number of sites within a cluster.
It allows for the calculation of the single-electron Green's function at arbitrary
values of the wave vector $\mathbf k$. This is a considerable improvement compared 
to standard Lanczos calculations for small clusters, where only a few $\mathbf k$ 
points are available.
The CPT has been successfully used to describe spectral properties of
the high-$T_C$ materials,\cite{Zacher00,Zacher02,Dahnken02} and has already
been extended to finite temperatures.\cite{Aichhorn03}

Recently a new method has been proposed which exploits a general variational
principle for the self-energy of a system of interacting fermions. This
self-energy-functional approach (SFA)\cite{Potthoff03} approximates the self
energy of the original system in the thermodynamic limit by the self energy
of an exactly solvable reference system with the same interaction part. The
self energy is varied by varying the single-particle parameters of the
reference system. Choosing the reference system to be a cluster of finite
size yields a non-perturbative and consistent cluster approach. It has
been shown\cite{Potthoff03_2} that within this framework the CPT as well as
the C-DMFT appear as special approaches depending on the number of additional
uncorrelated ("bath") sites taken into account: The optimum number of bath
sites is actually a free parameter which can be determined from the general
variational principle.
It has been pointed out \cite{Potthoff03_2} that at least for one-dimensional 
models a large cluster without bath sites has to be preferred.
The use of a reference system without bath sites represents a generalized CPT in 
which the single-particle parameters of the finite cluster are optimized 
according to the variational principle.
This "variational CPT" (V-CPT) has successfully been used in a recent study
for the investigation of the symmetry-broken antiferromagnetic phase of the 
two-dimensional Hubbard model.\cite{Dahnken03} 

So far a consistent formulation of the (variational) cluster-perturbation
approach could be achieved for lattice models with on-site interactions only.
The reason for this restriction is that within the SFA the reference system must be chosen
with the same interaction as the original model.
As detailed in \onlref{Potthoff03}, this ensures that functionals given by
the skeleton-diagram expansion are the same for both, the original and
the reference model.
In case of the EHM the interaction couples the different sites of the
lattice. Thus there is no reference system with the same interaction which
consists of decoupled subsystems of finite size.
The motivation of the present paper is therefore to extend the ideas of the CPT 
and V-CPT to the investigation of the EHM including nearest-neighbor Coulomb 
interaction. It is shown that a mean-field decoupling of the inter-cluster 
nearest-neighbor interaction yields a systematic and reliable cluster approach.

The paper is organized as follows: In \sect{sec:vcpt} we give a short description
of the V-CPT method, \sect{sec:decoupling} shows how to decouple clusters
in the case of the EHM. In \sect{sec:1d} and \sect{sec:2d} we present results
for one and two dimensions, respectively. The conclusions are given in 
\sect{sec:concl}.  

\section{Variational CPT}\label{sec:vcpt}

Let us consider a system of interacting fermions on a lattice with
Hamiltonian $H$, in general consisting of a single-particle part $H_0$ and an
interaction part $H_1$. The lattice is then divided into clusters, where it
is of crucial importance for the derivation of the method that those
clusters are connected by $H_0$ only. The Hamiltonian can then be written as
\begin{equation}\label{eq:ham_decoupled}
  H=\sum_\R\left[H_0^\intra(\R)+H_1(\R)\right]+\sum_{\R,\R^\prime}H_0^\inter(\R,\R^\prime),
\end{equation}
where $\R$ denotes the individual clusters, $H_0^\intra(\R)$ is the part of the
single-particle term that acts only inside a single cluster, $H_1(\R)$ is the
interaction part inside the cluster, and the
inter-cluster hopping is given by
\begin{equation}\label{eq:interclhop}
  H_0^\inter(\R,\R^\prime)=\sum_{a,b}T_{a,b}^{\R,\R^\prime}c_{\R,a}^\dagger
  c_{\R^\prime,b}^{\phantom{\dagger}},
\end{equation}
where the hopping matrix $T_{a,b}^{\R,\R^\prime}$ is non-zero only
for hopping processes across the cluster boundaries.
The indices $a$ and $b$ are general quantum numbers within a cluster,
e.g. position and spin index, and $c_{\R,a}^\dagger$ creates an electron with
quantum number $a$ in cluster $\R$.

The quantity of interest is the single particle Green's function
$G_{\R,a,\R^\prime,b}(\omega)=\langle\langle c_{\R,a}^{\phantom{\dagger}};
c_{\R^\prime,b}^\dagger\rangle\rangle_\omega$. Using translational
invariance at the level of the superlattice vector $\R$, the Green's function
becomes diagonal with respect to the wave vector $\mathbf Q$ from the reduced
Brillouin zone corresponding to the superlattice. The resulting Green's
function in reciprocal space is a matrix $\mathbf{G_Q}(\omega)$ with
elements $G_{{\mathbf Q},a,b}(\omega)$ and $a,b$ quantum numbers within a cluster.

Within the CPT approximation this Green's function $\mathbf{G_Q}(\omega)$
can be expressed in terms of Green's functions of the decoupled clusters
$\mathbf G^\prime(\omega)$, again matrices in the quantum numbers $a$ and
$b$, and the inter-cluster hopping $T_{a,b}^{\R,\R^\prime}$ by the expression
\begin{equation}\label{eq:cpt_gq}
  \mathbf{G_Q}(\omega)=\left[\mathbf
  G^\prime(\omega)^{-1}-\mathbf{T_Q}\right]^{-1}
\end{equation}
with the Fourier-transformed inter-cluster hopping
\begin{equation}\label{eq:cpt_tq}
  T_{\mathbf
  Q,a,b}=\frac{1}{L}\sum_{\R,\R^\prime}T_{a,b}^{\R,\R^\prime}e^{i\mathbf
  Q(\R-\R^\prime)}.
\end{equation}
For the details of the derivation of the CPT formulas we refer the
interested reader to Refs.\,\onlinecite{Senechal00,Senechal02} and references
therein. We want to mention that one can transform \eq{eq:cpt_gq} into a
Dyson-like equation
\begin{equation}\label{eq:cpt_dyson}
  \mathbf{G_Q}(\omega)=(\mathbf{G}_\mathbf{Q}^{(0)}(\omega)^{-1}-\mathbf\Sigma(\omega))^{-1},
\end{equation}
where $\mathbf{G}_\mathbf{Q}^{(0)}(\omega)$ is the free Green's function of
the infinite lattice, and
$\mathbf\Sigma(\omega)$ is the cluster self energy. In other words CPT 
consists of approximating the self energy of the infinite system by the self
energy of a cluster of finite size.
Note that CPT is based on the exact evaluation of small clusters
without any self-consistency procedure, and thus does not allow for the occurrence of
symmetry-broken phases. This restriction is overcome with the V-CPT
method.\cite{Potthoff03_2,Dahnken03} 

The observation underlying V-CPT is that the
perturbation term need not necessarily be restricted to 
the inter-cluster hopping term but can be any single-particle operator. For this
reason one has a certain amount of freedom for the partitioning of the
single-particle part of \eq{eq:ham_decoupled}. Namely, the Hamiltonian
\eq{eq:ham_decoupled} is invariant under the transformation
\begin{equation}\label{eq:transf}
  \begin{split}
  H_0^\intra(\R)&\to H_0^\intra(\R)+\mathcal{O}(\R)\\
  H_0^\inter(\R,\R^\prime)&\to H_0^\inter(\R,\R^\prime)
  -\delta_{\R,\R^\prime}\mathcal{O}(\R),
  \end{split}
\end{equation}
with an arbitrary intra-cluster single-particle operator
\begin{equation}\label{eq:add_onepartop}
  \mathcal{O}(\R)=\sum_{a,b}\Delta_{a,b}\, c_{\R,a}^\dagger
  c_{\R,b}^{\phantom{\dagger}},
\end{equation}
which can for instance be a fictitious symmetry-breaking field, thus allowing
for broken symmetry already on a finite system instead of only in the
thermodynamic limit.
In the non-interacting case, where \eq{eq:cpt_gq} is
exact,\cite{Senechal00,Senechal02} the result is independent 
of the transformation \eq{eq:transf}, but in the interacting case the result
depends on the particular choice. However, this dependence is not a
shortcoming of the method
but can rather be used for an optimization procedure.

The question of what choice for ${\mathbf \Delta}=\Delta_{a,b}$ will optimize the
results can be answered by the SFA.\cite{Potthoff03} It provides a way to
exactly evaluate the grand potential $\Omega[\mathbf \Sigma]$ as functional
of the self-energy $\mathbf\Sigma$ by restricting the domain of the functional to a
certain subspace $\mathcal S$ of trial self-energies. This subspace consists
of all $\mathbf\Sigma$ which are exact self-energies of the reference system
for different $\mathbf \Delta$. The reference system $H^\prime$ must have the
same interaction part as $H$ and must be exactly solvable for any $\mathbf \Delta$. 
Throughout the paper we use a cluster of finite size as reference system.

The cluster self-energy can be parametrized as $\mathbf\Sigma=\mathbf\Sigma(\mathbf
\Delta)$. Within the SFA, the optimal value of $\mathbf \Delta$ is determined
from the stationary point of the function\cite{Potthoff03}
\begin{align}\label{eq:sfa_omega}
  \Omega(\mathbf \Delta)&=\Omega^\prime(\mathbf \Delta)\nonumber\\
  &+T\sum_{\omega_n,\mathbf Q}\tr \ln\frac{-\mathbf
    1}{\mathbf{G}_\mathbf{Q}^{(0)}(i\omega_n)^{-1}-\mathbf\Sigma(\mathbf
    \Delta,i\omega_n)} \nonumber\\
  &-LT\sum_{\omega_n}\tr\ln(-\mathbf G^\prime(\mathbf \Delta,i\omega_n)),
\end{align}   
where $\Omega^\prime(\mathbf \Delta)$ is the grand potential
of the reference system. The frequency sum runs over discrete Matsubara
frequencies $i\omega_n$, $L$ 
is the number of clusters or $\mathbf Q$ points, respectively, $T$ gives the
temperature, and bold symbols denote
matrices in the cluster indices $a$ and $b$. Note that the fraction in the
second line in \eq{eq:sfa_omega} is the CPT Green's function. 
The single-particle  
parameters $\mathbf \Delta$ can include all single-particle parameters of the
original Hamiltonian or only part of it, as well as additional terms, e.g. a
fictitious staggered field. The more parameters are considered for the
optimization problem, the larger is the subspace $\mathcal S$ of trial self
energies. The actual choice and number of parameters depends on the problem under
consideration.
For more details of the derivation of the method see
\onlref{Dahnken03}. Note that the V-CPT method is superior to
the CPT approach, since within the concept of V-CPT one gets
the standard CPT formulas by setting the number of variational parameters
$\mathbf \Delta$ to zero. 

A necessary condition for the applicability of the method is that the
clusters are coupled by single-particle operators only. At this point it is easy
to see that a 
straightforward application of the method to the EHM where the clusters are also
coupled by Coulomb interactions is not possible. However, we will show
in \sect{sec:decoupling} how one can decouple the lattice into clusters
appropriate for the application of CPT even in the case of the EHM.

\section{Decoupling the clusters}\label{sec:decoupling}

We start from the Hamiltonian of the extended Hubbard model
\begin{align}
  H&=\sum_{ij,\sigma}T_{i,j}c_{i\sigma}^\dagger c_{j\sigma}^{\phantom{\dagger}} + U\sum_i
  n_{i\uparrow}n_{i\downarrow}\nonumber\\
  &+V\sum_{\langle ij\rangle}n_in_j -\mu\sum_in_i,\label{eq:ehm}
\end{align}
where $i,j$ indicate the position in the lattice, and for convenience we use a
constant value $V_{i,j}\equiv V$ for all nearest-neighbor bonds. 
According to
\eq{eq:ham_decoupled} we decouple the lattice into clusters yielding
\begin{align}
    H&=\sum_\R\left[H_0^\intra (\R)+H_{\rm U}^\intra (\R)+H_{\rm V}^\intra (\R)\right]\nonumber\\
    &+\sum_{\R,\R^\prime}
    \left[H_0^\inter
    (\R,\R^\prime)+H_{\rm V}^\inter (\R,\R^\prime)\right],\label{eq:ehm_decoupled}
\end{align}
where the first row includes only terms of a single cluster and the 
second row couples different clusters. By comparing the
second row with the corresponding term in
\eq{eq:ham_decoupled} one can see that the term causing problems in
the case of the EHM is the interaction term
\begin{equation}\label{eq:ehm_int_term}
  H_V^\inter(\R,\R^\prime)=V\sum_{[ij]} n_{\R i}n_{\R^\prime j},
\end{equation}
which is of two-particle type. The symbol $[ij]$ indicates that the sum
runs only over bonds connecting nearest neighbors in different
clusters. For nearest-neighbor interactions this means that the indices
in $[ij]$ must belong to the cluster boundaries of two adjacent clusters. For
the application of the method derived in \sect{sec:vcpt} the 
coupling term must be of single-particle type, which can be achieved by a
mean-field decoupling of the interaction term
\eq{eq:ehm_int_term}. Hence we get
\begin{align}
  H_{\rm V,MF}^\inter (\R,\R^\prime )&=V\sum_{[ij]} \left[n_{\R i}\langle n_{\R^\prime
      j}\rangle + \langle  n_{\R i}\rangle n_{\R^\prime j}\right]\nonumber\\
  &-V\sum_{[ij]}\langle n_{\R
  i}\rangle\langle n_{\R^\prime j}\rangle.
\end{align}
Due to the translational invariance with respect to the superlattice vector $\R$,
the mean-field parameters $\langle n_{\R i}\rangle$ and $\langle
n_{\R^\prime j}\rangle$ are independent of $\R$ and $\R^\prime$ and will be
denoted by $\lambda_i$ and $\lambda_j$, respectively. With these abbreviations we get
\begin{align}
  \sum_{\R,\R^\prime}
  &H_{\rm V,MF}^\inter (\R,\R^\prime) =\nonumber\\
  &=V \sum_{\R,\R^\prime}
  \sum_{[ij]}\left[n_{\R i}\lambda_j+n_{\R^\prime j}\lambda_i-\lambda_i\lambda_j\right]\nonumber\\
  &=V\sum_{\R}\sum_{[ij]}\left[n_{\R i}\lambda_j+n_{\R
  j}\lambda_i-\lambda_i\lambda_j\right]\nonumber\\ 
  &=\sum_{\R}H_{\rm V,MF}^\inter (\R).\label{eq:ehm_mf_approx}
\end{align}  
The double sum over $\R$ and $\R^\prime$ reduces to a single sum, because for 
fixed values of $\R$, $i$, and $j$ only one term of the sum over $\R^\prime$ contributes
due to the fact that two-site interactions couple at most two different
clusters. One has to be careful in order to avoid
double counting of the bonds $[ij]$. For instance, for a one-dimensional cluster of 
length $N$, \eq{eq:ehm_mf_approx} reduces to
\begin{equation}
  V\sum_{\R}\left[n_{\R 1}\lambda_N + n_{\R N}\lambda_1-\lambda_1\lambda_N\right],
\end{equation}
because the only decoupled bond connects sites $1$ and $N$ of different clusters.

By this mean-field decoupling, two parameters $\lambda_i$ are introduced for
each decoupled bond, e.g. $\lambda_1$ and $\lambda_N$ in one dimension, and
in general all these parameters $\lambda_i$ are independent of each
other. But as we will see below, the number of mean-field parameters
$\lambda_i$ can be strongly reduced in special cases. 

The decoupled interaction \eq{eq:ehm_mf_approx} is of single-particle type and
can be included in the intra-cluster hopping term $H_0^\intra(\R)$, leading to
a modified intra-cluster single-particle term
\begin{equation}
  \tilde H_0^\intra(\R,\lambda_i) = H_0^\intra(\R)+H_{\rm V,MF}^\inter (\R,\lambda_i),
\end{equation}
where we explicitly denoted the dependence on the parameters $\lambda_i$. 
After mean-field decoupling we finally get the Hamiltonian
\begin{align}
  H_{\rm MF}(\lambda_i)&=\sum_\R\left[\tilde H_0^\intra (\R,\lambda_i)+H_{\rm U}^\intra
    (\R)+H_{\rm V}^\intra (\R)\right]\nonumber\\ 
    &+\sum_{\R,\R^\prime}
    H_0^\inter(\R,\R^\prime),\label{eq:ehm_mf_decoupled}
\end{align}
for which the method described in \sect{sec:vcpt} is applicable.

From the decoupling of the clusters we have got additional parameters $\lambda_i$
which are external parameters to the Hamiltonian
\eq{eq:ehm_mf_decoupled} and have to be determined in a proper way. For
this purpose we propose two different procedures:

(i) One can get the
parameters from a self-consistent calculation on an isolated
cluster. That means that one starts with a certain guess for the $\lambda_i$,
which are the expectation values of the electron densities on sites $i$. 
Then the ground-state wave function of an isolated cluster is calculated, giving
new values for the $\lambda_i$. In this step open boundary conditions (obc) are
used in order to be consistent with the obc necessary for the calculation of
the cluster Green's function in Eqs.~(\ref{eq:cpt_gq})
and~(\ref{eq:sfa_omega}). These new values
$\lambda_i$ serve as parameters in the Hamiltonian for the next determination of the ground
state, and the whole procedure is iterated until convergence of the
$\lambda_i$ is achieved. This
procedure may work quite well for the EHM in the case of a first order phase
transition between a disordered and an ordered phase, because (due to an avoided
level crossing) the transition point, i.e. the critical Coulomb interaction
$V_c$, is almost independent of the cluster size.\cite{Sengupta02} For
second order phase transitions we expect that this method will not give
satisfactory results, because here we face a discrepancy between the
parameters calculated on the isolated cluster and the parameters that would
give the optimal result in the thermodynamic limit.
    
(ii) The shortcoming in the case of second order phase transitions can be
overcome in the following way: As we show in App.~\ref{app:mf}, the
self-consistent calculation of mean-field parameters is equivalent to the
minimization of the free energy $F$.
Since the relation $\Omega=F-\mu N$ holds at $T=0$, this minimization can be
done at the same time as the optimization 
of the single-particle parameters $\mathbf \Delta$ in the SFA formalism,
and we can use \eq{eq:sfa_omega} for the
determination of the parameters $\lambda_i$, too. Note that all quantities in 
\eq{eq:sfa_omega} which depend on the single-particle parameters
$\mathbf \Delta$ are dependent on the mean-field parameters
$\lambda_i$ as well. 
To keep the calculations simple we consider only half-filled
systems, where it is sufficient to use only two different values for the
$\lambda_i$, namely $\lambda_A=1-\delta$ and $\lambda_B=1+\delta$ on sublattices $A$
and $B$, respectively. Under this assumption we have only one mean-field parameter
$\delta$, and the grand potential is $\Omega = \Omega(\mathbf
\Delta,\delta)$. The general procedure is now, that for each value of
$\delta$ the stationary point with respect to $\mathbf \Delta$ has to be
found as required by the SFA formalism, yielding a function
$\Omega=\Omega(\delta)$. By finding the minimum of this function one can determine
the optimal value for $\delta$.

Conceptually, the latter method (ii) of determining the mean-field parameters 
is superior to the procedure (i) described first as it uses information on the
Green's function in the thermodynamic limit for the calculation of $\delta$. 
However, one has to keep in mind that for each choice of $\delta$ the Green's 
function $\mathbf G^\prime(\omega)$ of the isolated cluster has to be
calculated many times
to evaluate \eq{eq:sfa_omega} which is much more time consuming than the
self-consistency procedure on the isolated cluster.

\section{One dimension}\label{sec:1d}

The Hamiltonian of the one-dimensional EHM is given by
\begin{align}
  H=&-t\sum_{i,\sigma}\left(c_{i,\sigma}^\dagger
  c_{i+1,\sigma}^{\phantom{\dagger}}+\text{H.c.}\right) 
  + U\sum_in_{i\uparrow}n_{i\downarrow}\nonumber\\
  &+V\sum_in_in_{i+1}-\mu\sum_in_i.\label{eq:1dehm}
\end{align}
Throughout the paper we set $t$ as the unit of energy. Although this model was
studied intensively in recent years, the ground state phase
diagram is still under some
discussion.\cite{Sengupta02,Hirsch84,Nakamura99,Nakamura00,Tsuchiizu02,Jeckelmann02,Tsuchiizu03}
We use this model as a testing ground 
for our method, because many results are available for comparison. The
chemical potential is $\mu=U/2+2V$ due to particle-hole symmetry at half-filling.

\begin{figure}
  \centering
  \includegraphics[width=0.35\textwidth]{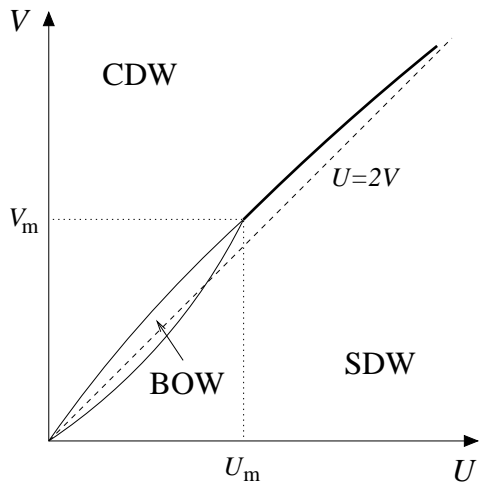}
  \caption{\label{fig:shem_phasediagram}%
  Schematic phase diagram of the one-dimensional EHM taken from
  Refs.~\onlinecite{Sengupta02,Tsuchiizu02,Tsuchiizu03}. The thick line marks
  the first order phase transition, and the dashed line marks $U=2V$.}
\end{figure}

In one dimension at half filling, the EHM shows an involved phase diagram including spin
density wave (SDW), charge density wave (CDW), and bond order wave (BOW)
phases as shown schematically in \fig{fig:shem_phasediagram}. There is good
agreement on the existence of the latter phase, but its 
extension in the $U$-$V$-plane has not yet been clarified in
detail.\cite{Sengupta02,Jeckelmann02} Moreover, a multi-critical point occurs
at some critical value $U_m$, where the phase transition into the ordered CDW
phase changes from a second order transition at lower values to a first
order transition at higher values. Recent QMC studies\cite{Sengupta02} gave
$U_m\approx4.7$ for this point in good agreement with $g$-ology
\cite{Tsuchiizu02,Tsuchiizu03} and bosonization results,\cite{Voit92} whereas
DMRG gives a somewhat lower value $U_m\approx 3.7$.\cite{Jeckelmann02}

\subsection{First order phase transition}\label{subsec:1du8}

For a first test of our method we studied the one-dimensional EHM at $U=8$,
which is well above the multi-critical point. The phase transition is then
of first order without any BOW phase between SDW and CDW phases. As reference
system $H^\prime$ according to \sect{sec:vcpt}, we used decoupled clusters of
different lengths consisting of $N_c=8$, 10, and 12 sites, respectively. For the
determination of the mean-field parameter $\delta$ we used the method (ii)
described in \sect{sec:decoupling}, where $\delta$ is calculated from the
minimum 
of the free energy of the system. For the SFA optimization of the
single-particle parameters, we had to choose a set $\mathbf \Delta$ of
parameters which are varied in the optimization procedure. In order to
minimize the number of relevant parameters we used results of a recent study
of the Hubbard model.\cite{Potthoff03_2} There it has been shown that at $U=8$ the variation
of the hopping in the cluster yields only minor changes that can be 
neglected, and that open boundary conditions have to be used. Moreover, for
the one-dimensional Hubbard model it has been pointed out that the use 
of a fictitious staggered magnetic field as a variational parameter gives a
stationary point of the grand potential only for vanishing field, yielding a
paramagnetic phase without long-range magnetic order. Since it can be assumed
that these results are also valid in a similar way for the EHM, we did not
use the hopping in the cluster or a staggered magnetic field in the
optimization procedure. Here we studied charge-ordering effects and therefore
we used as variational parameter a staggered field coupled to the charge
densities given by \eq{eq:add_onepartop} with 
\begin{equation}\label{eq:staggeredfield}
  \Delta_{a,b}=\varepsilon\delta_{a,b}e^{i\mathbf{Q}\R_a},
\end{equation}
where $\mathbf Q=\pi$ is the wave vector of staggered ordering and
$\varepsilon$ is the staggered-field strength.
\begin{figure}[t]
  \centering
  \includegraphics[width=0.45\textwidth]{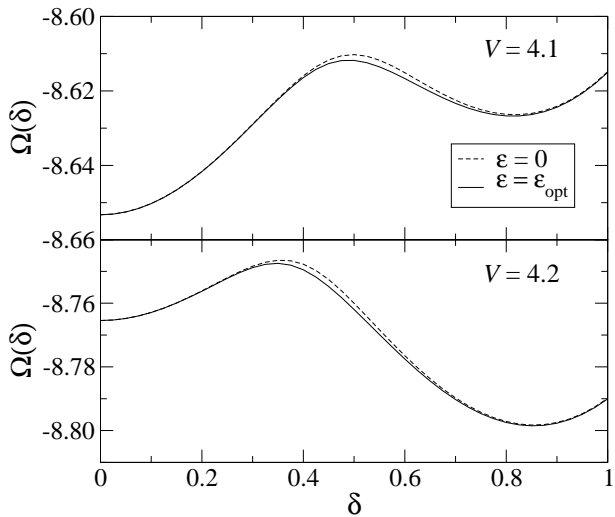}
  \caption{\label{fig:gp_1du8}%
    Grand potential $\Omega$ as a function of the mean-field parameter
    $\delta$ at $U=8$ calculated on a cluster with $N_c=8$ sites as
    reference system. Upper panel: $V=4.1$. Lower panel: $V=4.2$. Solid
    lines: With optimization of a staggered field. Dashed lines: Without
    optimization of a staggered field.}
\end{figure}
The grand potential obtained in this way is shown in \fig{fig:gp_1du8} at
two values of the inter-site Coulomb interaction. For comparison, calculations
without optimization of the staggered field
are shown as dashed lines in \fig{fig:gp_1du8}. As one
can see, the optimization gives only minor changes to $\Omega(\delta)$. The
optimal staggered-field strengths in these calculations varied between $\varepsilon_{\rm
  opt}=0.0$ at $\delta=0.0$ and $\varepsilon_{\rm  opt}\approx 0.05$ at $\delta=1.0$
at both values of $V$.

From the shape of $\Omega(\delta)$ one can directly infer the order of the
transition. If three minima 
occur at $\delta=0$ and $\delta=\pm\delta_{\rm CDW}$, it is of first order,
whereas it is of second order if $\Omega(\delta)$ has only two minima at
$\delta=\pm\delta_{\rm CDW}$ and a maximum at $\delta=0$. As on can easily
see in \fig{fig:gp_1du8}, we have clear evidence for a first order phase
transition at $U=8$ with an SDW minimum at $\delta=0.0$ and two degenerate CDW
minima at $\delta=\pm \delta_{\rm CDW}$. At $V=4.1$ the SDW phase is
realized, $\Omega(0)<\Omega(\delta_{\rm CDW})$,
whereas at $V=4.2$ we have $\Omega(0)>\Omega(\delta_{\rm CDW})$ and the CDW
phase is the stable one. Thus we can
state that the critical value $V_c$ for the phase transition is located
between $V=4.1$ and $V=4.2$.

\begin{figure}[t]
  \centering
  \includegraphics[width=0.45\textwidth]{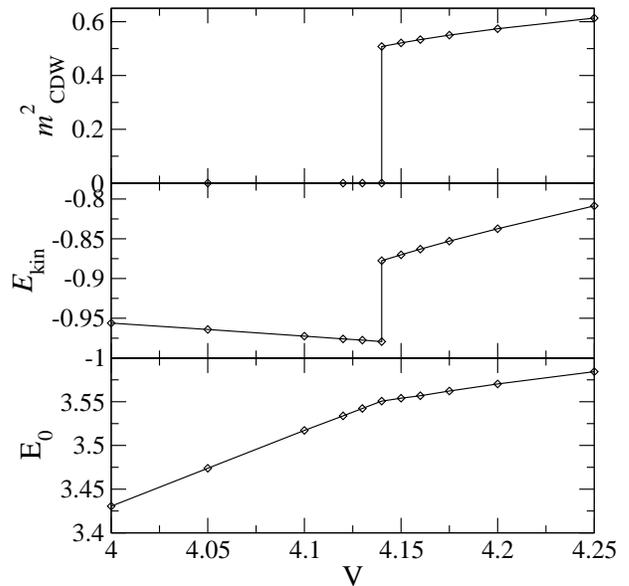}
  \caption{\label{fig:res_1du8}%
    Ground state energy $E_0$, kinetic energy $E_{\rm kin}$, and order
    parameter $m_{\rm CDW}^2$ of the one-dimensional EHM at $U=8$ after
    finite size scaling. Lines are guides to the eye only.}
\end{figure}

For a more accurate determination of the phase boundary $V_c$, we have
calculated the grand potential at several values of $V$ and cluster sizes
$N_c=8$, 10, and 12. In addition to the grand potential and the ground state
energy $E_0=\Omega+\mu N_e$ with $N_e$ the number
of electrons in the system, we calculated the order parameter
\begin{equation}\label{eq:cdworderpar}
  m_{\rm CDW} = \frac{1}{N_c}\sum_j\left(n_j-\langle n\rangle\right)e^{i\mathbf
  Q\R_j},
\end{equation}
where $\mathbf Q=\pi$, $N_c$ is the number of cluster sites, and the kinetic
energy $E_{\rm kin}$. Both properties can be extracted from the spectral 
function $A_{\R,\mathbf r,\R^\prime,\mathbf r^\prime,\sigma}(\omega)=(-1/\pi){\rm Im} \langle\langle
c_{\R,\mathbf r,\sigma}^{\phantom{\dagger}};c_{\R^\prime,\mathbf
  r^\prime,\sigma}^\dagger\rangle\rangle^{\rm (ret)}_\omega$. The kinetic
energy is obtained after
Fourier-transformation\cite{Senechal00,Senechal02,Dahnken03} 
\begin{equation}\label{eq:akw_ft}
  A(\mathbf k,\omega)=\frac{1}{LN_c}\sum_{\R,\R^\prime}\sum_{\mathbf
  r,\mathbf r^\prime}e^{i\mathbf k(\R+\mathbf r-\R^\prime-\mathbf
  r^\prime)}A_{\R,\mathbf r,\R^\prime,\mathbf r^\prime,\sigma}(\omega)
\end{equation}
by
\begin{equation}\label{eq:ekin}
  E_{\rm kin}=\frac{2}{L}\sum_{\mathbf
  k}\int_{-\infty}^0\!\!\!{\rm d}\omega\,\varepsilon(\mathbf k)
  A(\mathbf k,\omega),
\end{equation}
with $\varepsilon(\mathbf k)$ the dispersion of the non-interacting
system. Within our approach it is necessary to use the Lehmann representation
for the cluster Green's function with small but finite Lorentzian broadening $\sigma$.
Whereas the grand potential \eq{eq:sfa_omega} shows only minor dependence on this
broadening, the dependence of
the order parameter and the kinetic energy is considerably larger and one has
to do an extrapolation to $\sigma=0$.\cite{Senechal02}
Although the formalism applies to the thermodynamic limit, results show a
finite size dependence due to the finite size of the clusters serving as
reference system. We found that the order parameter exhibits the strongest
finite-size effects, which were of the order $m^2_{{\rm CDW},N_c=10}/m^2_{{\rm
    CDW},N_c=12}\approx 1.02$ at all values of $V$. Finite-size scaling to
$N_c=\infty$ is easily done and the results for the ground state energy extracted 
from the minimum of the grand potential, the kinetic energy and the order
parameter are shown in \fig{fig:res_1du8}. Our results should
be compared to Fig.\,10 of \onlref{Sengupta02} which shows excellent
quantitative agreement with a deviation of less than 2\,\% for the
calculated quantities at all values of $V$. From our calculations we get
$V_c= 4.140(5)$, again in agreement with the previous studies 
Refs.\,\onlinecite{Sengupta02,Jeckelmann02}. 
  
\begin{figure*}
  \centering
  \includegraphics[width=0.45\textwidth,height=0.23\textheight]{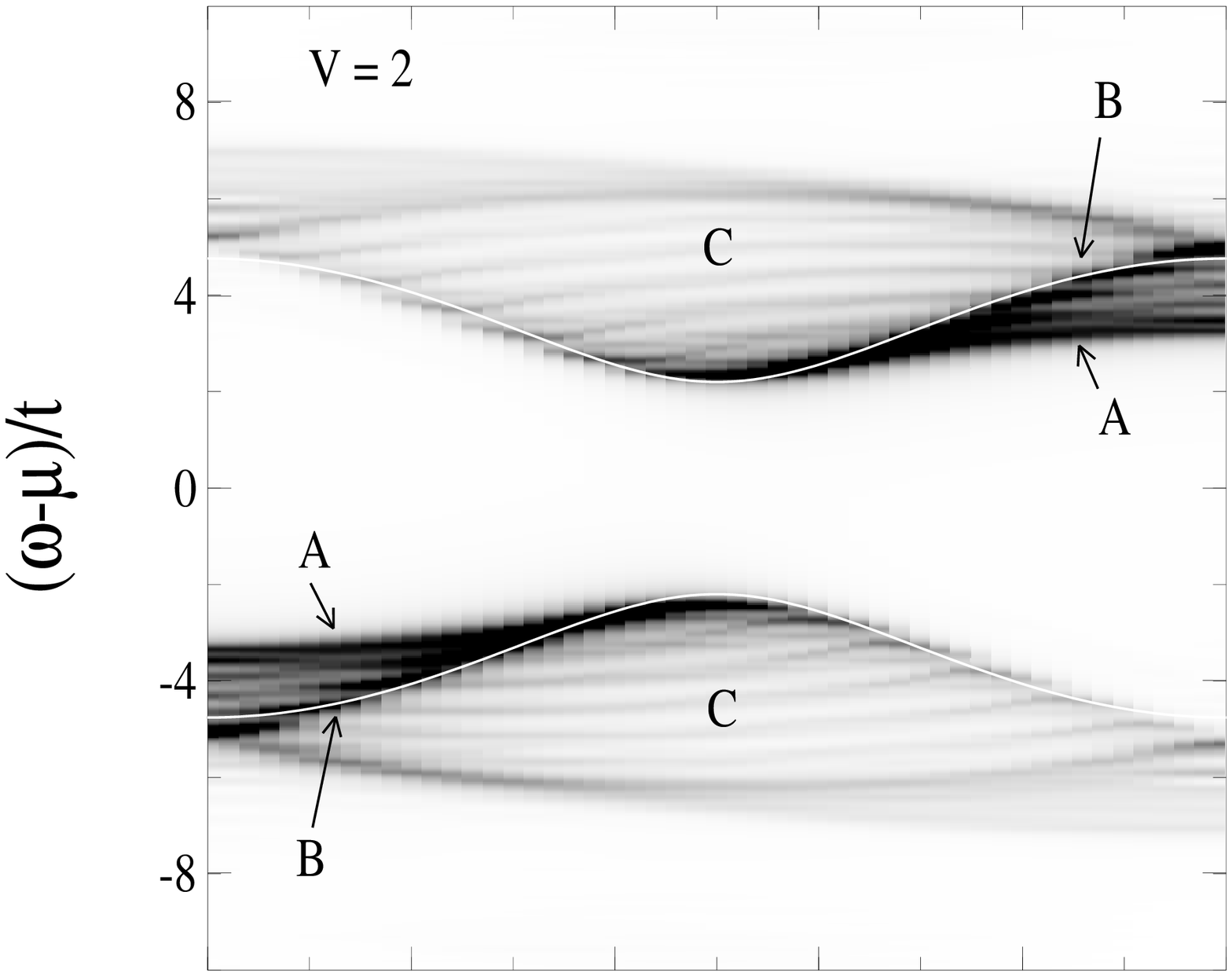}
  \includegraphics[width=0.45\textwidth,height=0.23\textheight]{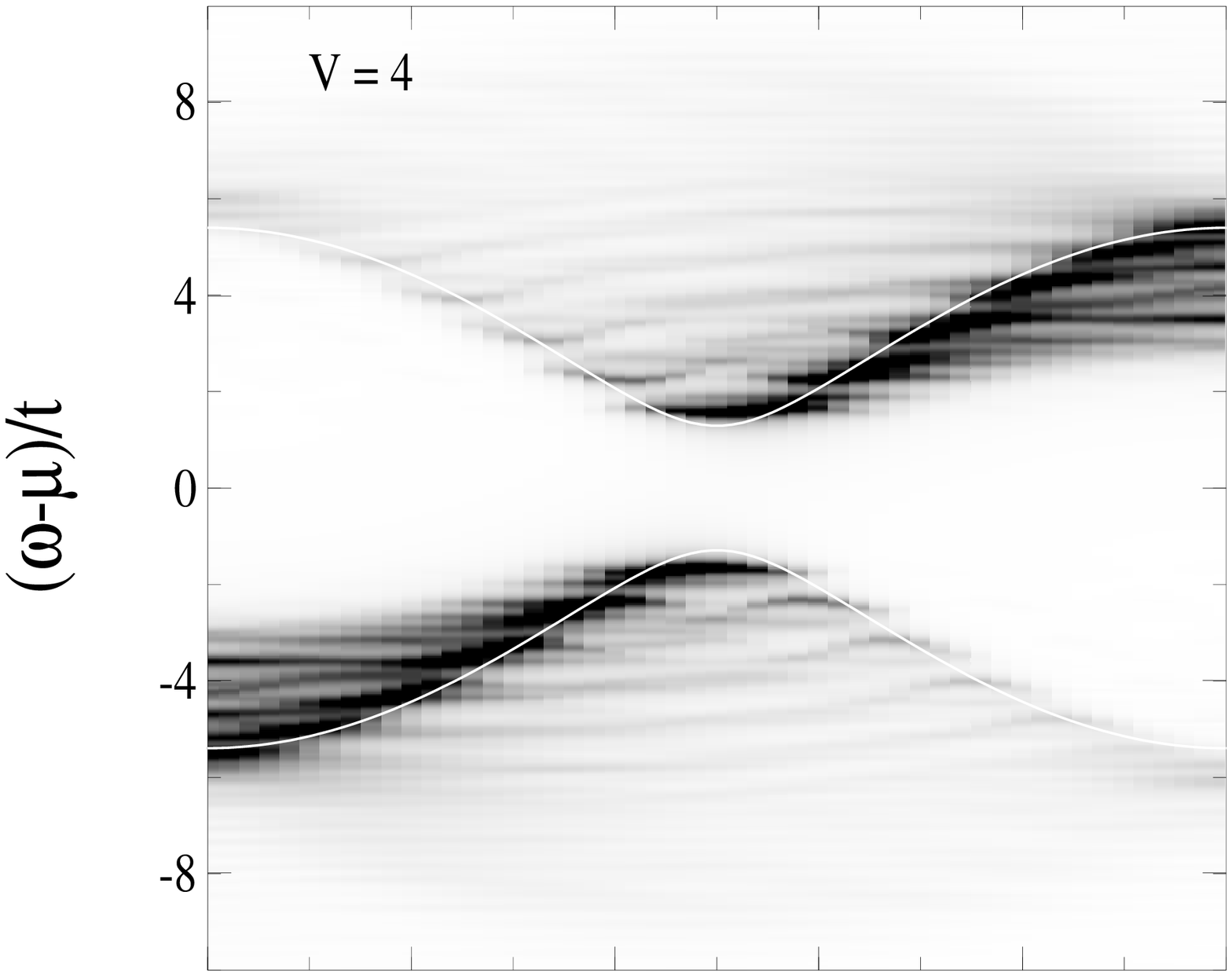}
  \includegraphics[width=0.45\textwidth,height=0.28\textheight]{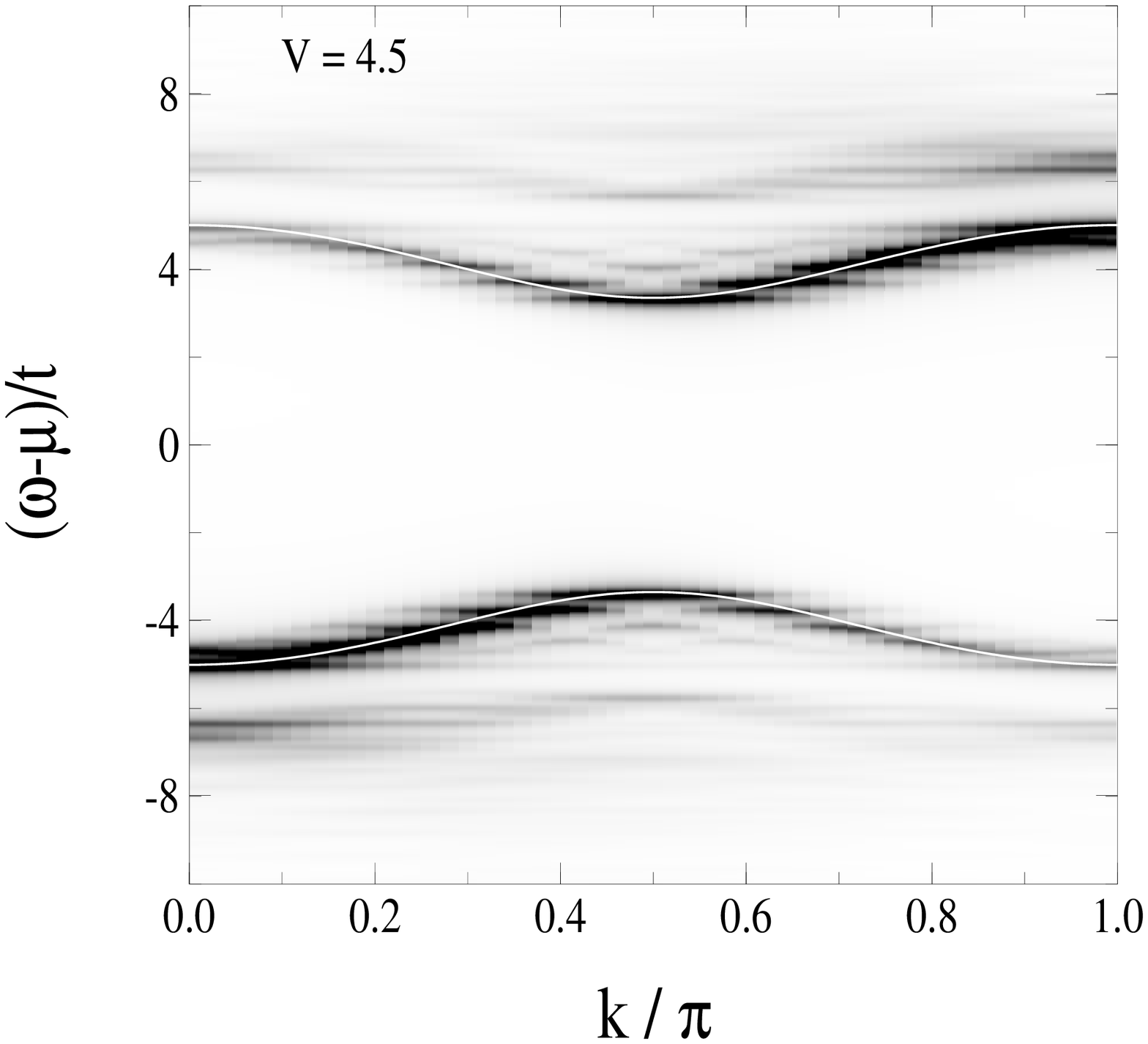}
  \includegraphics[width=0.45\textwidth,height=0.28\textheight]{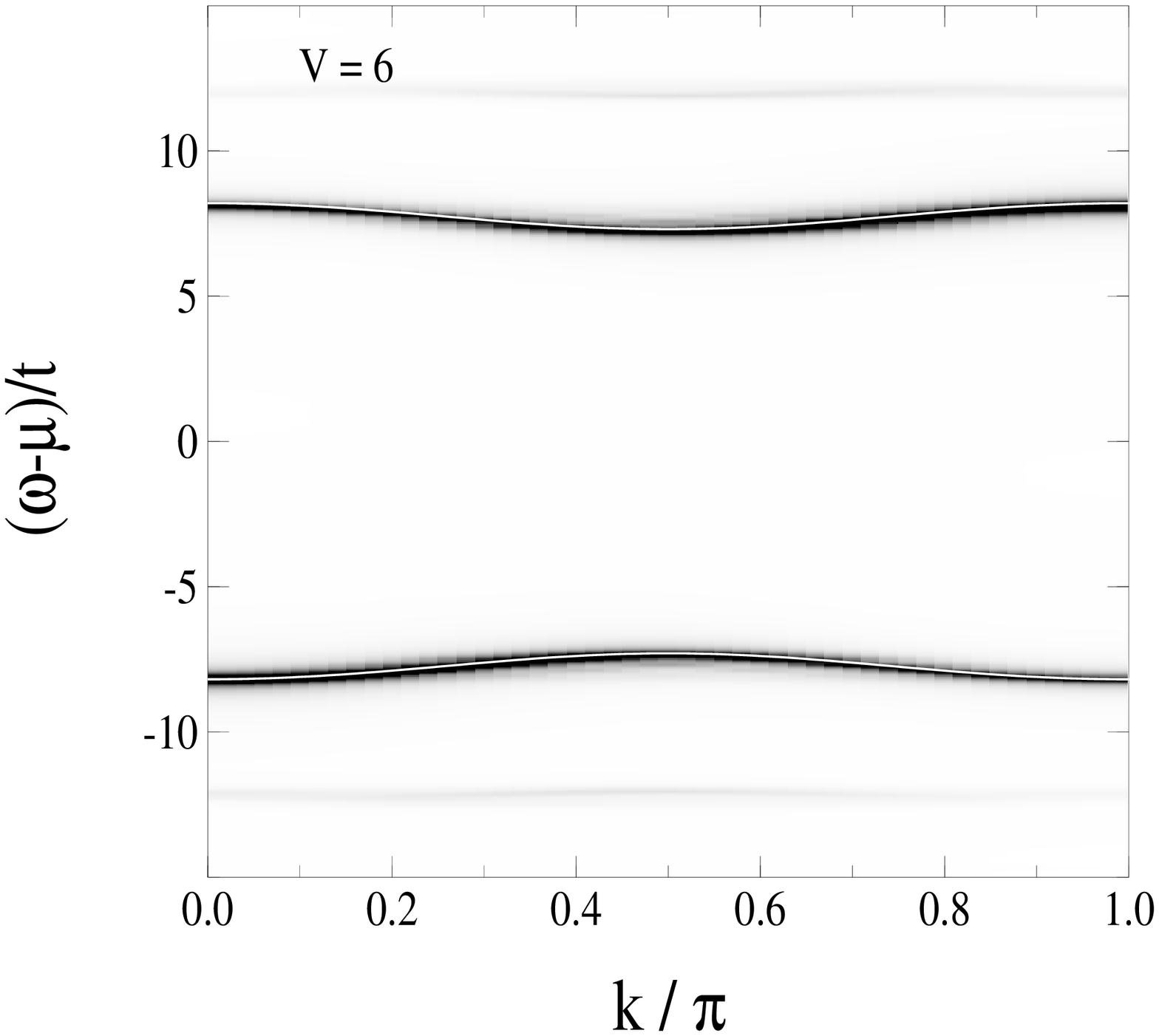}
  \caption{\label{fig:akw_1du8}%
    Density plot of the spectral function $A(\mathbf k,\omega)$ of the one-dimensional EHM at
    $U=8$, calculated on a cluster of size $N_c=12$ with Lorentzian broadening
    $\sigma=0.1$. Darker regions represent larger spectral weight. Coulomb
    interaction $V$ as indicated in the plots. White lines are fits to a Hartree-Fock SDW/CDW
    dispersion (see text).}
\end{figure*}

In order to provide a complete picture of the method we also performed calculations
with mean-field parameters obtained by a self-consistent procedure on an
isolated cluster, see method (i) in \sect{sec:decoupling}. For instance for $N_c=12$ and
$V=4.1$ one finds self-consistent solutions for $\delta=0$ and for
$\delta_{\rm SC}=0.832$, which 
differs only slightly from the value extracted from the grand potential,
$\delta_{\rm CDW}=0.822$. For this reason the calculation of the ground-state
energy, kinetic energy, and order parameter using $\delta_{\rm SC}$ instead
of $\delta_{\rm CDW}$ gives practically the same results as in \fig{fig:res_1du8}.
In the present case it is therefore sufficient to 
calculate the mean-field parameter from an isolated cluster which is much
faster than finding the minimum of the grand potential.

Whereas the properties we have shown so far are well known for the
one-dimensional EHM, we additionally calculated for the first time the spectral function by
\eq{eq:akw_ft} for arbitrary wave vector $\mathbf k$. In \fig{fig:akw_1du8}
results are shown at $U=8$ and selected values of $V$ with a reference
system consisting of $N_c=12$ cluster sites, and the mean-field parameter $\delta$ calculated
self-consistently by method (i), see \sect{sec:decoupling}. We want to
mention that the 'striped' structure, particularly visible in the regions marked by 'C' in
\fig{fig:akw_1du8}, occurs because the decoupling into clusters breaks the
translational invariance of the system.

The spectral function at $V=2.0$ is very similar to the spectral function of
the Hubbard model ($V=0$)\cite{Senechal00,Senechal02} with splitting of the
low-energy band into a spinon and an holon band, which are marked in
\fig{fig:akw_1du8}
by 'A' and 'B', respectively. This similarity could have already been 
expected based on the full Hartree-Fock solution---decoupling of all
interaction terms in the Hamiltonian---where one has no dependence
on $V$ at all in the SDW phase. But this simple picture holds only away from
the transition point $V_c$ as can be seen in \fig{fig:akw_1du8} in the
plot at $V=4.0$. At this point, in the vicinity of the phase transition  
$V_c= 4.14$, the gap is considerably smaller than at
$V=2.0$, a clear deviation from the Hartree-Fock prediction. This
indicates that charge fluctuations become very important in this regime, which 
are completely neglected by the Hartree-Fock approximation, but are taken into
account on the length scale of the cluster in our approach. But although we
found this deviation, one can still see residuals of the splitting of the
low-energy band, a signature for spin-charge separation. For this reason we
infer that spin-charge separation is present up to the transition point.
\begin{table}
  \centering
  \caption{\label{tab:fit_1du8}%
     Fitted values for the hopping matrix element $t_{\rm fit}$, gap
     $\Delta_{\rm fit}$, and gap $\Delta_{\rm HF}$ of the full
     Hartree-Fock approximation at $U=8$.} 
  \begin{tabular}{|c|c|c|c|}
    \hline
     & $t_{\rm fit}$ & $\Delta_{\rm fit}$ & $\Delta_{\rm HF}$\\
     \hline
    $\quad V=0.0\quad$ & $\quad 1.93\quad$ & $\quad 2.24\quad$ & $\quad 3.75\quad$\\
    $V=2.0$ & 2.11 & 2.20 & 3.75\\
    $V=4.0$ & 2.62 & 1.29 & 3.75\\
    $V=4.5$ & 1.86 & 3.35 & 4.80 \\
    $V=6.0$ & 1.86 & 7.29 & 7.88\\
    \hline
  \end{tabular}
\end{table}
The white lines in \fig{fig:akw_1du8} correspond to fits of the holon branch
to a Hartree-Fock dispersion $E(\mathbf
k)=\pm\sqrt{\Delta^2+\varepsilon(\mathbf k)^2}$. The fitted values for the
hopping matrix element $t_{\rm fit}$ and the gap $\Delta_{\rm fit}$ are
denoted in \tabl{tab:fit_1du8} where we included the values at $V=0$ for 
completeness. One finds that the gap $\Delta_{\rm fit}$ is almost constant
from $V=0$ to $V=2$ and, as mentioned above, considerably decreases 
near the the phase transition ($V=4$). The hopping matrix element $t_{\rm
  fit}$ shows the opposite behavior and increases when approaching the transition point
from below. This is due to the fact that in the vicinity of $V_c$,
doubly-occupied and singly-occupied sites become close in energy, which
enhances the movement of the electrons. The actual value of the matrix
element $t_{\rm fit}$ is very large compared to the original value $t=1$ in
the Hamiltonian. A fit to the spinon band would give a smaller value closer
to $t=1$, but whereas fitting to the holon band is consistent over the
whole range of momentum vectors $\mathbf k$, the spinon band is only present
for $\mathbf k<\pi/2$ for $\omega-\mu<0$ (and $\mathbf k>\pi/2$ for
$\omega-\mu>0$, respectively).

The spectral function in the CDW phase shows a qualitatively different
behavior. At $V=4.5$ we found a gap considerably larger than in the SDW phase, and this gap
increases very fast with increasing $V$, as can be seen in the plot at
$V=6$. Moreover, no evidence for spin-charge separation can be seen in the
spectral functions. By comparing the fitted value $\Delta_{\rm fit}$ with
the Hartree-Fock solution $\Delta_{\rm HF}$, one can see that the
agreement at $V=4.5$ is better than at $V=4$, and that it becomes
still better with increasing $V$. For this reason we conclude that
charge fluctuations which are neglected in the Hartree-Fock approximation
play a minor role in the CDW phase. 

\subsection{Second order phase transition}

So far all calculations were done at $U=8$, where the system shows a first
order phase transition. In the following, we study the EHM at $U=3$,
where the model exhibits a second order transition into the charge ordered CDW
phase.\cite{Sengupta02,Jeckelmann02} In this paper we do not consider the BOW, since
it has been argued that the SDW-BOW transition is of Kosterlitz-Thouless
type.\cite{Nakamura00} For an analysis of this type of transition the
available cluster sizes are far too small and do not allow a clear distinction between
SDW and BOW phase.

\begin{figure}
  \centering
  \includegraphics[width=0.45\textwidth]{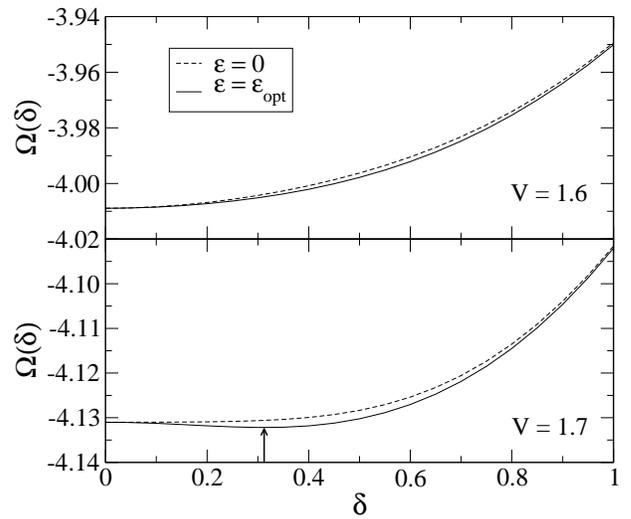}
  \caption{\label{fig:gp_1du3}%
    Grand potential $\Omega$ as function of the mean-field parameter
    $\delta$ at $U=3$ calculated on a cluster with $N_c=8$ sites as reference
    system. Upper panel: $V=1.6$. Lower panel: $V=1.7$. Solid lines: With
    optimization of a staggered field. Dashed lines: Without optimization of a
    staggered field. The arrow marks the CDW minimum at $V=1.7$.}
\end{figure}

We calculate the grand
potential $\Omega(\delta)$ in the same way as in \sect{subsec:1du8} in order
to determine $\delta$. The result of a calculation on a cluster consisting of
$N_c=8$ sites is shown in \fig{fig:gp_1du3}.
One can easily see a striking difference between the grand potential at
$U=8$, \fig{fig:gp_1du8}, and at $U=3$. In the latter case there is only a
single minimum. It is located at $\delta=0$ for $V<V_c$. With increasing $V$
the curve for $\Omega(\delta)$ becomes flatter in the region 
around $\delta=0$ and finally two degenerate CDW minima occur at
$\delta=\pm\delta_{\rm CDW}$ for $V>V_c$. Note that here $\delta$ changes
continuously when crossing $V_c$, whereas it shows a discontinuity in the case
of a first order phase transition.

We find that now it is indeed important to use a staggered field,
\eq{eq:staggeredfield}, as a variational SFA parameter. In \fig{fig:gp_1du3},
results are shown with such an optimization (solid lines) and without (dashed
lines). Whereas at $V=1.6$ both calculations show only the SDW minimum at
$\delta=0$, they differ at $V=1.7$ where the system should already be in the
charge-ordered phase.\cite{Sengupta02,Nakamura99,Nakamura00,Jeckelmann02}
Without optimization of the staggered field, we would still find the SDW
minimum at $\delta=0$, but with optimization the minimum shows up for a
finite value of $\delta=\pm 0.31$ characteristic for the CDW
phase.

\begin{figure}
  \centering
  \includegraphics[width=0.45\textwidth]{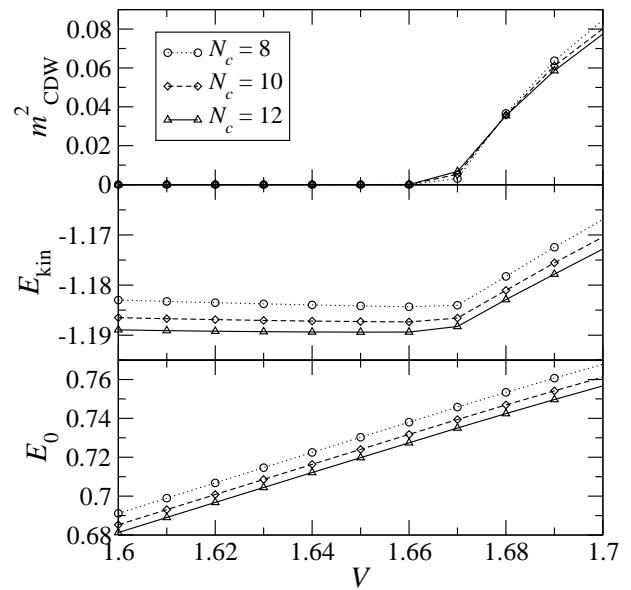}
  \caption{\label{fig:res_1du3}%
    Ground state energy $E_0$, kinetic energy $E_{\rm kin}$, and order
    parameter $m_{\rm CDW}^2$ of the one-dimensional EHM at $U=3$ for
    cluster sizes $N_c=8$ (dotted), $N_c=10$ (dashed), and $N_c=12$ (solid line).}
\end{figure}
\begin{figure}
  \centering
  \includegraphics[width=0.45\textwidth,height=0.23\textheight]{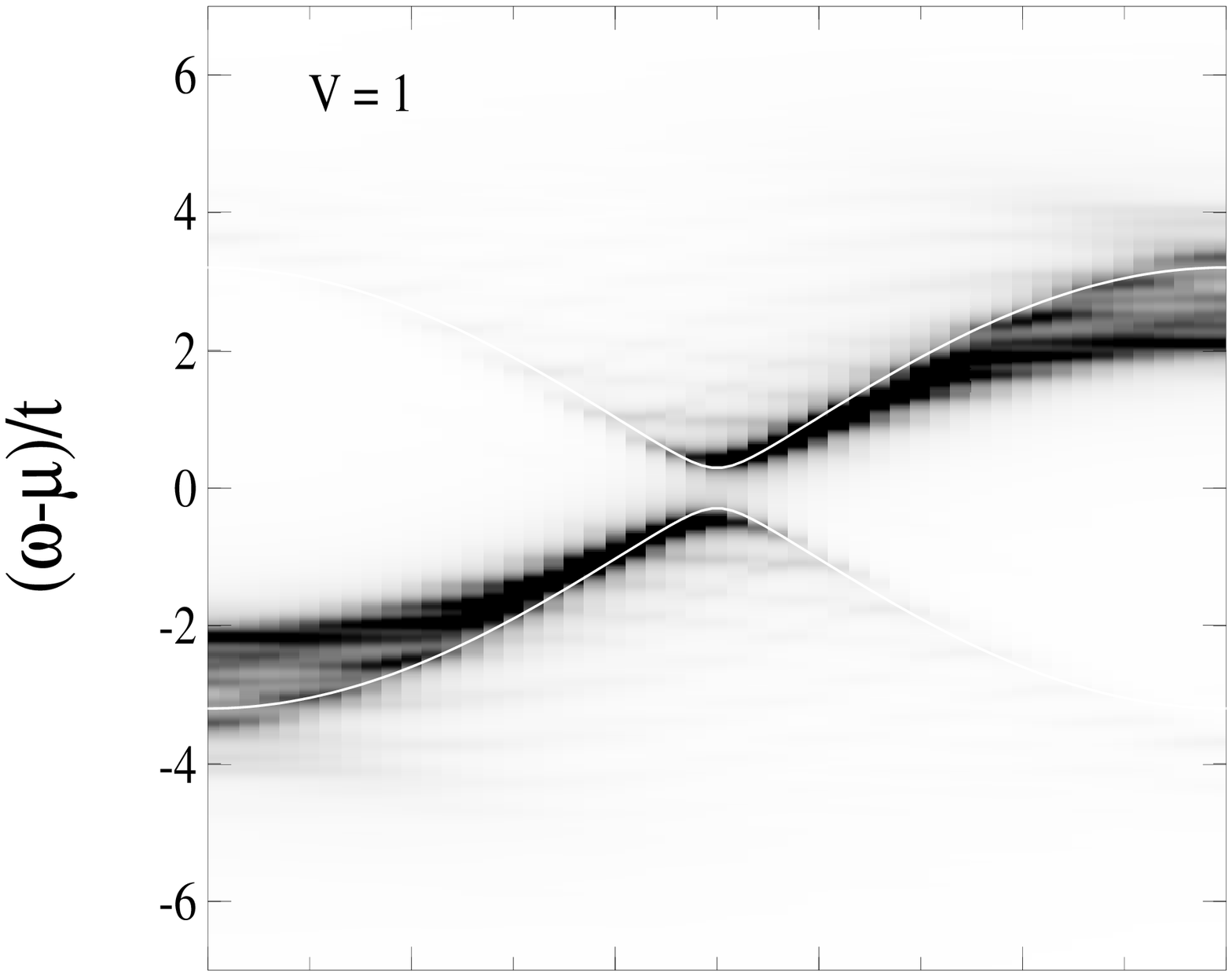}
  \includegraphics[width=0.45\textwidth,height=0.23\textheight]{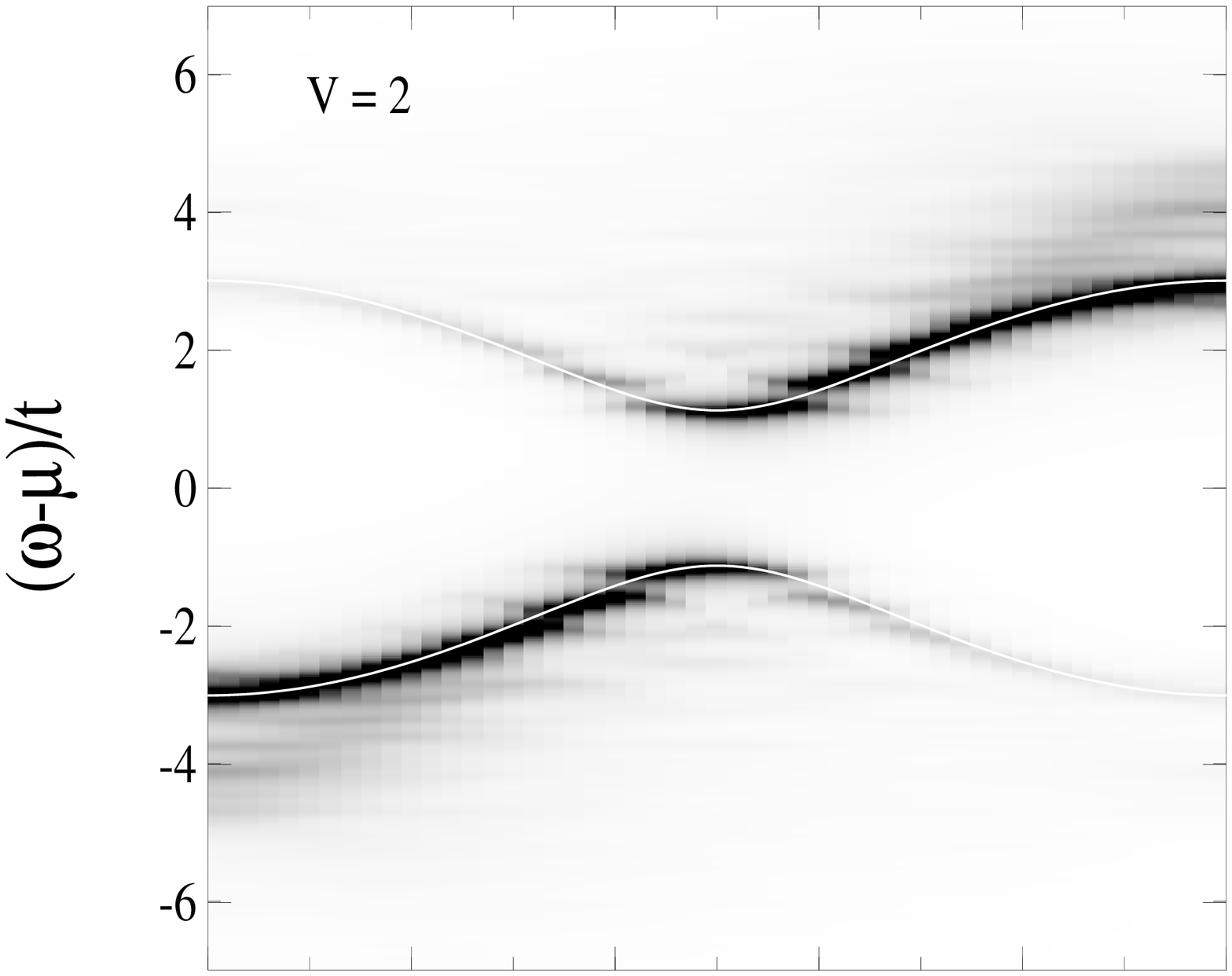}
  \includegraphics[width=0.45\textwidth,height=0.28\textheight]{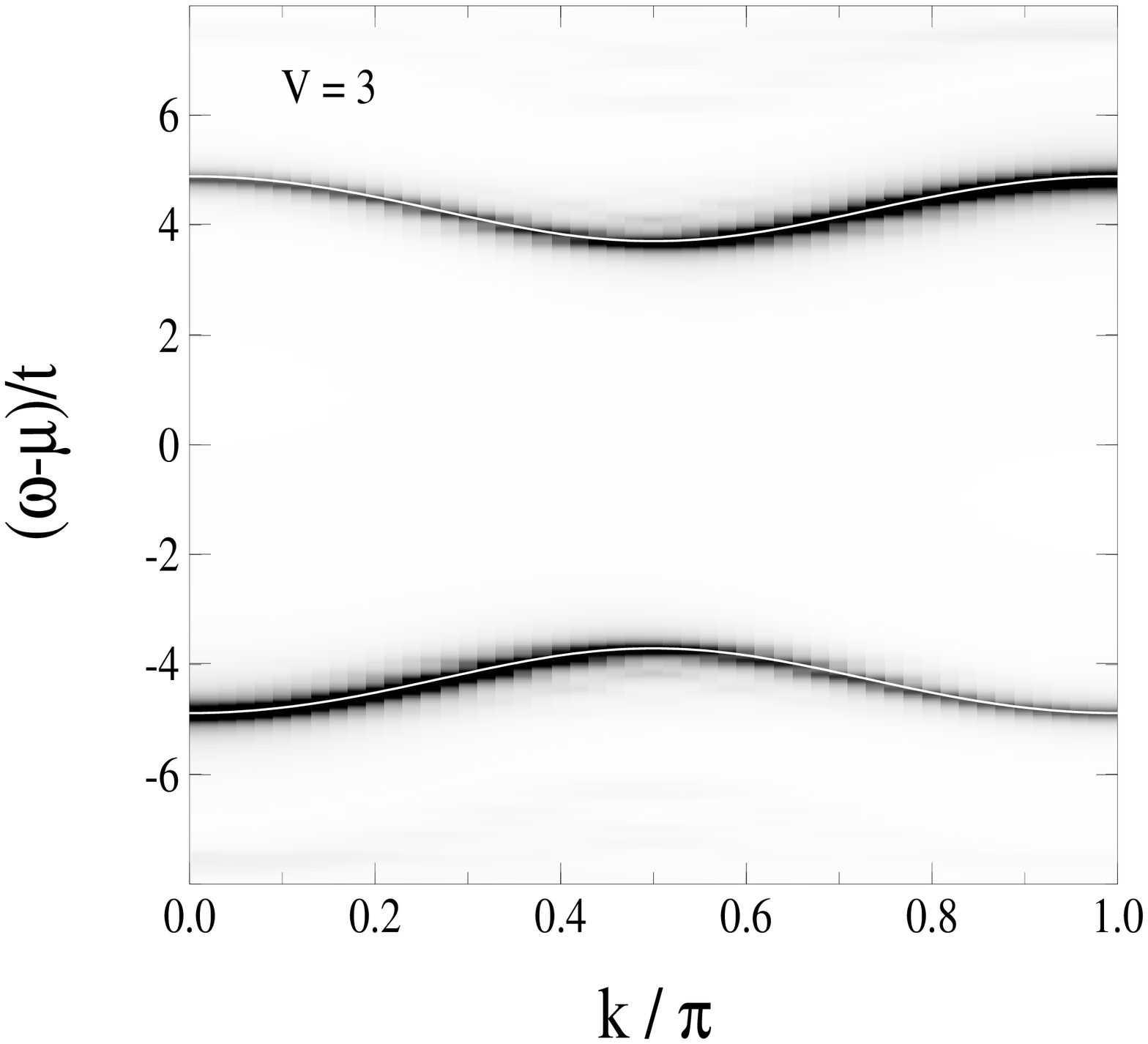}
  \caption{\label{fig:akw_1du3}%
    Density plot of the spectral function $A(\mathbf k,\omega)$ of the one-dimensional EHM at
    $U=3$ calculated on a cluster of size $N_c=12$ with Lorentzian broadening
    $\sigma=0.1$. Darker regions represent larger spectral weight. From top to bottom:
    $V=1.0$, $2.0$, $3.0$. White lines are fits to a Hartree-Fock SDW/CDW
    dispersion (see text).}
\end{figure}

For the determination of the critical value $V_c$, we calculated the ground
state energy $E_0$, kinetic energy $E_{\rm kin}$, and the order parameter
$m_{\rm CDW}$ at several values of $V$ shown in \fig{fig:res_1du3}. We performed
no finite-size scaling like in \sect{subsec:1du8} since we found that here
the cluster sizes are too small for a systematic scaling. Different
from \fig{fig:res_1du8}, $E_{\rm kin}$, $m_{\rm CDW}$, and the slope of the
ground state energy are continuous
across the transition point as required for a second order transition. From
the kinetic energy and the order parameter calculated on a cluster of size
$N_c=12$, we extract a critical value of $V_c=1.665(5)$, which is in good
agreement with the critical value $V_c\approx 1.65$ obtained by QMC\cite{Sengupta02} and
diagonalization methods,\cite{Nakamura99,Nakamura00} and with $V_c=1.64(1)$ from
DMRG calculations.\cite{Jeckelmann02}
The slight difference is likely due to remaining finite-size
effects. Moreover we made use of a single variational parameter only, namely
the staggered field \eq{eq:staggeredfield}, and it can be expected that
including more single-particle parameters in the SFA optimization procedure
would give even more accurate results. 

We would like to point out that in the present case of a second order phase transition, the
most accurate way of calculating the mean-field parameter $\delta$ is to find
the minimum in the grand potential including SFA optimization of
single-particle parameters. Calculations on a cluster of size $N_c=12$ showed
that without optimization the critical value would be
$V_c= 1.685(5)$. Compared to $V_c=1.665(5)$ this is further away from the
values obtained by other methods as given above. Calculations with $\delta$ 
obtained self-consistently on an isolated cluster are insufficient. In this
case one would get $V_c=1.735(5)$ for the $N_c=12$ cluster. This means that
for a second order phase transition $\delta$ should be determined by
minimizing the grand potential, whereas for first order transitions the self
consistent determination was sufficient.

\begin{table}
  \centering
  \caption{\label{tab:fit_1du3}%
     Fitted values for the hopping matrix element $t_{\rm fit}$, gap
     $\Delta_{\rm fit}$, and $\Delta_{\rm HF}$ within the Hartree-Fock
     approximation at $U=3$.}  
  \begin{tabular}{|c|c|c|c|}
    \hline
     & $t_{\rm fit}$ & $\Delta_{\rm fit}$ & $\Delta_{\rm HF}$\\
     \hline
    $\quad V=0\quad$ & $\quad 1.38\quad$ & $\quad 0.29\quad$ & $\quad 0.93\quad$\\
    $V=1$ & 1.59 & 0.29 & 0.93\\
    $V=2$ & 1.40 & 1.13 & 2.12\\
    $V=3$ & 1.59 & 3.71 & 4.28\\
    \hline
  \end{tabular}
\end{table}

The spectral function $A({\mathbf k},\omega)$ at $V=1.0$, $2.0$, and $3.0$,
which has not been calculated previously, is
depicted in \fig{fig:akw_1du3}. We found that the spectral function at $V=1.0$
shows only minor differences to the spectral function of the Hubbard model
($V=0$). The white lines in \fig{fig:akw_1du3} are fits to a Hartree-Fock
SDW/CDW 
dispersion. The parameters $t_{\rm fit}$ and $\Delta_{\rm fit}$ can be read
off from \tabl{tab:fit_1du3}. In the SDW phase at $V=0$ and $V=1.0$, the gap
$\Delta_{\rm fit}$ is constant. Similar to the case $U=8$ the agreement
between $\Delta_{\rm fit}$ and $\Delta_{\rm HF}$ is better in the CDW phase
than in the SDW phase. The hopping parameter $t_{\rm fit}$
increases when approaching the phase transition from below, similar to
\tabl{tab:fit_1du8}, but the fitted values for $t_{\rm fit}$ are
considerably smaller than in the case $U=8$.

\section{Two dimensions}\label{sec:2d}

\begin{figure}
  \centering
  \includegraphics[width=0.25\textwidth]{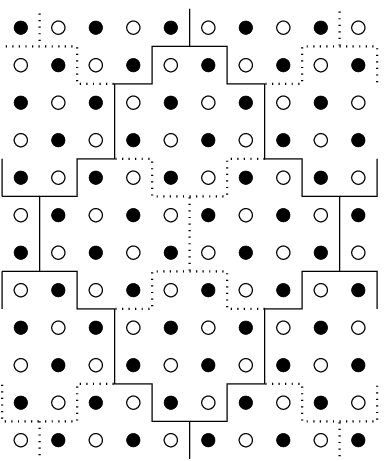}\\[0.4cm]
  \includegraphics[width=0.45\textwidth]{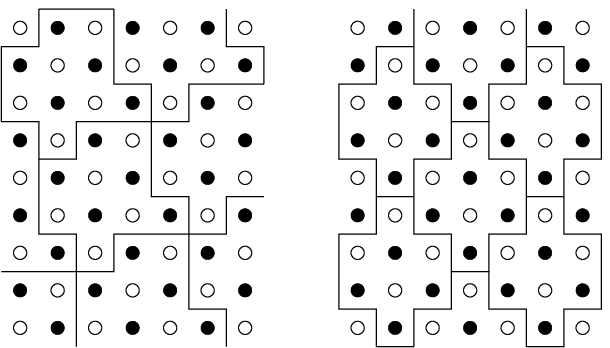}
  \caption{\label{fig:clusters}%
    Possible tilings of the two dimensional square lattice into clusters that
    allow for staggered ordering: $N_c=8$ (bottom right), $N_c=10$
    (bottom left), super cluster with $N_c=48$ (top).}
\end{figure}

The two-dimensional Hubbard model is one of the most intensely discussed
models for strongly-correlated electron systems, especially in the context of
high-temperature superconductivity. But different from the one-dimensional
case, where many sophisticated methods have been used to investigate the extended
Hubbard model as described in \sect{sec:1d}, only few studies have been done for
the two-dimensional EHM. One reason for this is that many modern methods such as
DMRG or fermionic loop-update QMC are difficult to apply to more than one spatial
dimension. However, within our present approach, the extension to two
dimensions is straightforward.

The two-dimensional EHM is defined by the Hamiltonian
\begin{align}
  H=&-t\sum_{\langle ij\rangle,\sigma}\left(c_{i,\sigma}^\dagger
    c_{j,\sigma}^{\phantom{\dagger}}+\text{H.c.}\right)
  + U\sum_in_{i\uparrow}n_{i\downarrow}\nonumber\\
  &+V\sum_{\langle ij\rangle} n_in_j-\mu\sum_in_i,\label{eq:2dehm}
\end{align}
where $\langle ij\rangle$ connects nearest neighbors and the chemical
potential is $\mu=U/2+4V$ at half filling. Early QMC studies\cite{Zhang89} showed
that this model has a SDW-CDW transition similar to the one-dimensional
case with transition point $V_c\approx U/4$. But due to numerical
difficulties it was impossible to determine the exact position and the order
of the phase transition. Calculations within the Hartree-Fock 
approximation\cite{Chattop97}
showed two stable phases for the Hamiltonian \eq{eq:2dehm} at half filling,
the SDW and CDW phase, separated by a phase boundary at $V_c=U/4$. 

For the application of the method presented in \sect{sec:vcpt}, the
two-dimensional square lattice has to be decoupled into clusters of finite
size. Three possible tilings with different numbers of cluster sites $N_c$
are shown in \fig{fig:clusters}. Some care has 
to be taken concerning the staggered ordering. Whereas for clusters with
$N_c=8$ and $N_c=10$ shown in \fig{fig:clusters}, the staggered ordering
indicated by open and full circles is commensurate over the cluster
boundaries, a straightforward decoupling into clusters of size $N_c=12$ 
is not possible. As one can easily see, a super cluster with $N_c=48$
consisting of four 
$N_c=12$ clusters has to be constructed in order to take into account the
staggered ordering correctly. The Green's function of the super cluster can
be calculated by switching off the hopping processes that connect the
single $N_c=12$ clusters, in other words on bonds across the dotted lines in
\fig{fig:clusters}. This gives a block-diagonal Hamiltonian which can be treated by the
Lanczos algorithm. The switched off hopping processes are then incorporated again
perturbatively, that means by including the corresponding hopping terms in
the matrix $T_{a,b}^{\R,\R^\prime}$ in \eq{eq:cpt_tq}. Note that here the
vectors $\R$ and $\R^\prime$ denote the super clusters and not the single
$N_c=12$ clusters.
Of course there are many other possible tilings like the $4\times3$ cluster used in
Refs.\,\onlinecite{Senechal00,Senechal02,Senechal03}, but also in that case a
super cluster of $N_c=24$ has to be used.

\begin{figure}
  \centering
  \includegraphics[width=0.45\textwidth]{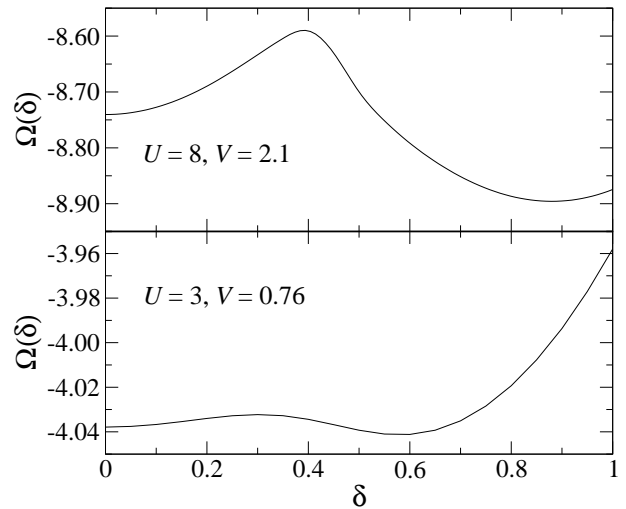}
  \caption{\label{fig:gp_2d}%
    Grand potential calculated on a cluster of size $N_c=8$ at
    $U=8,V=2.1$ (upper panel) and $U=3, V=0.76$ (lower panel).}
\end{figure}

We start the analysis of the two-dimensional EHM with the determination of
the order of the phase transition. For this purpose we use the $N_c=8$
cluster shown in \fig{fig:clusters} and calculate the grand potential
$\Omega(\delta)$ in the vicinity of the transition point at $U=8.0$ and
$U=3.0$ as described in method(ii) in \sect{sec:decoupling}. Here we did not
use a staggered field as variational parameter, because it does not change
the qualitative shape of $\Omega(\delta)$ (see Figs.~\ref{fig:gp_1du8} and
\ref{fig:gp_1du3}) and is therefore not necessary for the 
determination of the order of the transition. The result of this 
calculation is shown in \fig{fig:gp_2d}. At
both values of $U$ we found three minima, located at $\delta=0$ and
$\delta=\pm \delta_{\rm CDW}$. This indicates a first order phase transition,
different from the one-dimensional EHM, where at $U=3.0$ the transition is
of second order. We checked that this different behavior is not likely to be
a finite size effect due to the 
small linear dimension of the two-dimensional $N_c=8$ cluster by calculating
$\Omega(\delta)$ for the one-dimensional model with $N_c=4$ which still shows clear
evidence of a second-order phase transition at $U=3$.

The fact that the system shows first order transitions at both $U=8.0$ and $U=3.0$
simplifies the subsequent calculations. As discussed in the previous section,
one gets good results in the case of a first order transition by using a mean-field
parameter $\delta$ determined self consistently on an isolated cluster, as
described in method (i) in \sect{sec:decoupling}. This
procedure is much faster than the calculation of the grand potential for many
values of $\delta$, which makes it possible to use the $N_c=48$ super cluster
shown in \fig{fig:clusters}. We want to mention at this point that the
calculation of the grand potential for the two-dimensional system is much
more time consuming than for one dimension because of the larger number of
$\mathbf{Q}$ points required in \eq{eq:sfa_omega}. For one dimension $L\approx 40$ is
sufficient for convergence, whereas $L\approx 500$ is necessary
for two dimensions. Nevertheless it is of crucial importance to use a cluster
as large as possible, because the ratio of bonds treated exactly to
mean-field decoupled bonds increases with increasing cluster size, especially
pronounced for the two-dimensional square lattice. 
After having determined the mean-field parameter $\delta$ for the CDW phase
self consistently, we also performed an SFA optimization of a staggered field,
\eq{eq:staggeredfield}. 

\begin{figure}
  \centering
  \includegraphics[width=0.5\textwidth]{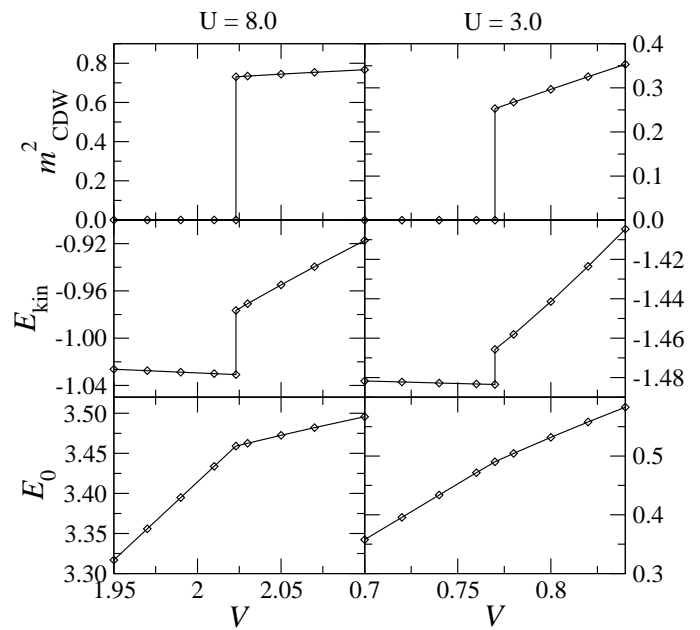}
  \caption{\label{fig:res_2d}%
    Ground state energy $E_0$, kinetic energy $E_{\rm kin}$, and order
    parameter $m^2_{\rm CDW}$ of the 2D EHM at $U=8.0$ (left) and $U=3.0$
    (right). Calculations were done on a $N_c=48$ super cluster.}
\end{figure}

A few more words have to be said about calculations in the SDW phase
($\delta=0$). Recent studies\cite{Dahnken03} of the pure Hubbard model revealed
that it is important to 
take into account the long-range magnetic order for the accurate description
of salient features of the system. This can be achieved by using a staggered
magnetic field as variational parameter, given by
\begin{equation}\label{eq:staggeredfield_magn}
  \Delta_{a,b}=h\delta_{a,b}z_\sigma e^{i\mathbf Q\R_a},
\end{equation}
with $z_\sigma=\pm 1$ for spin projection $\sigma=\uparrow,\downarrow$, and
$h$ the strength of the field.
Additionally it was argued that due to
the connection of the hopping parameter $t$ and the magnetic exchange
constant $J$, results could be further improved by letting the hopping in the
clusters be of strength $t^\prime$ and optimizing the staggered
magnetic field and $t^\prime$ simultaneously. Therefore we use
\begin{equation}
  \Delta_{a,b}=h\delta_{a,b}z_\sigma e^{i\mathbf Q\R_a}-\tau\delta_{\langle
  ab\rangle^\prime}, 
\end{equation}
where the symbol $\delta_{\langle ab\rangle^\prime}$ is equal to one for
nearest-neighbor bonds inside the cluster and zero otherwise.
The field strength $h$ and $\tau=t^\prime -t$ are the variational parameters in the
optimization procedure.

To sum up, the following steps are performed in the analysis using the
$N_c=48$ super cluster: (i) First we
determine the mean-field parameter $\delta_{\rm CDW}$ in the CDW phase
self-consistently on an isolated cluster and (ii) use a staggered field
\eq{eq:staggeredfield} for an SFA optimization procedure. (iii) In the SDW
phase ($\delta=0$) the staggered magnetic field \eq{eq:staggeredfield_magn}
and the intra-cluster hopping $t^\prime$ are 
optimized simultaneously. (iv) After determination of the SFA variational
parameters we calculate the quantities we are interested in.

The results for the ground state energy, kinetic energy, and order parameter
are shown in \fig{fig:res_2d}. At both $U=8.0$ and $U=3.0$, the behavior of
a first order transition can be seen, where the change in the slope of $E_0$
is much stronger at $U=8.0$ than at $U=3.0$. This change at $U=8.0$
is even more pronounced than for the one-dimensional model at $U=8.0$.
From \fig{fig:res_2d} we can extract the critical value
$V_c$ of the phase transition by fitting $E_0$ to a straight line in the
vicinity of the transition point, and for the $N_c=48$ super cluster we find
$V_c= 2.023(1)$ at $U=8.0$, and 
$V_c= 0.770(3)$ at $U=3.0$. These values of $V_c$ are much
closer to the Hartree-Fock result $V_c=U/4$ than for one
dimension. Within our approach we cannot clarify whether this is an intrinsic
feature of the two-dimensional model or it is an artifact of the
approximation due to the larger number of mean-field decoupled bonds.

The SFA variational parameters in the SDW phase near the phase transition
point are found to be almost
independent of the interaction $V$. At $U=8$, the optimization resulted in 
$t^\prime\approx 1.1$ for the intra-cluster hopping and $h\approx 0.14$
for the staggered magnetic field. The optimization of just one single parameter
leads to $t^\prime \approx 1.03|_{h=0}$ and $h\approx
0.12|_{t^\prime=t}$, and the value of $\Omega$ also differs significantly from
the value obtained by the simultaneous optimization of $t^\prime$ and
$h$. This means that due to the strong connection between the 
magnetic ordering and the hopping matrix element it is important to optimize
$t^\prime$ and $h$ simultaneously in order to get the best
approximation for the physics in the thermodynamic limit.
In the charge ordered phase the dependence
of the variational parameter, \eq{eq:staggeredfield}, on the interaction $V$ is larger with
$\varepsilon=0.08$ at $V=2.01$ and $\varepsilon=0.22$ at $V=2.1$.
A similar behavior can be found at $U=3$: In the SDW phase the variational
parameters $t^\prime\approx 1.61$ and $h\approx 0.15$ are almost independent
of $V$. In the CDW phase we get $\varepsilon=-0.03$ at $V=0.76$ and
$\varepsilon=-0.18$ at $V=0.84$.

Whereas the application of the 
magnetic staggered field exhibits the symmetry $h\to-h$, this
is not the case for the staggered field \eq{eq:staggeredfield}, because the
symmetry is already broken by the mean-field decoupling. 
We found no stationary point of $\Omega$ for finite $h$ in the CDW phases.

\begin{figure}
  \centering
  \includegraphics[width=0.45\textwidth,height=0.24\textheight]{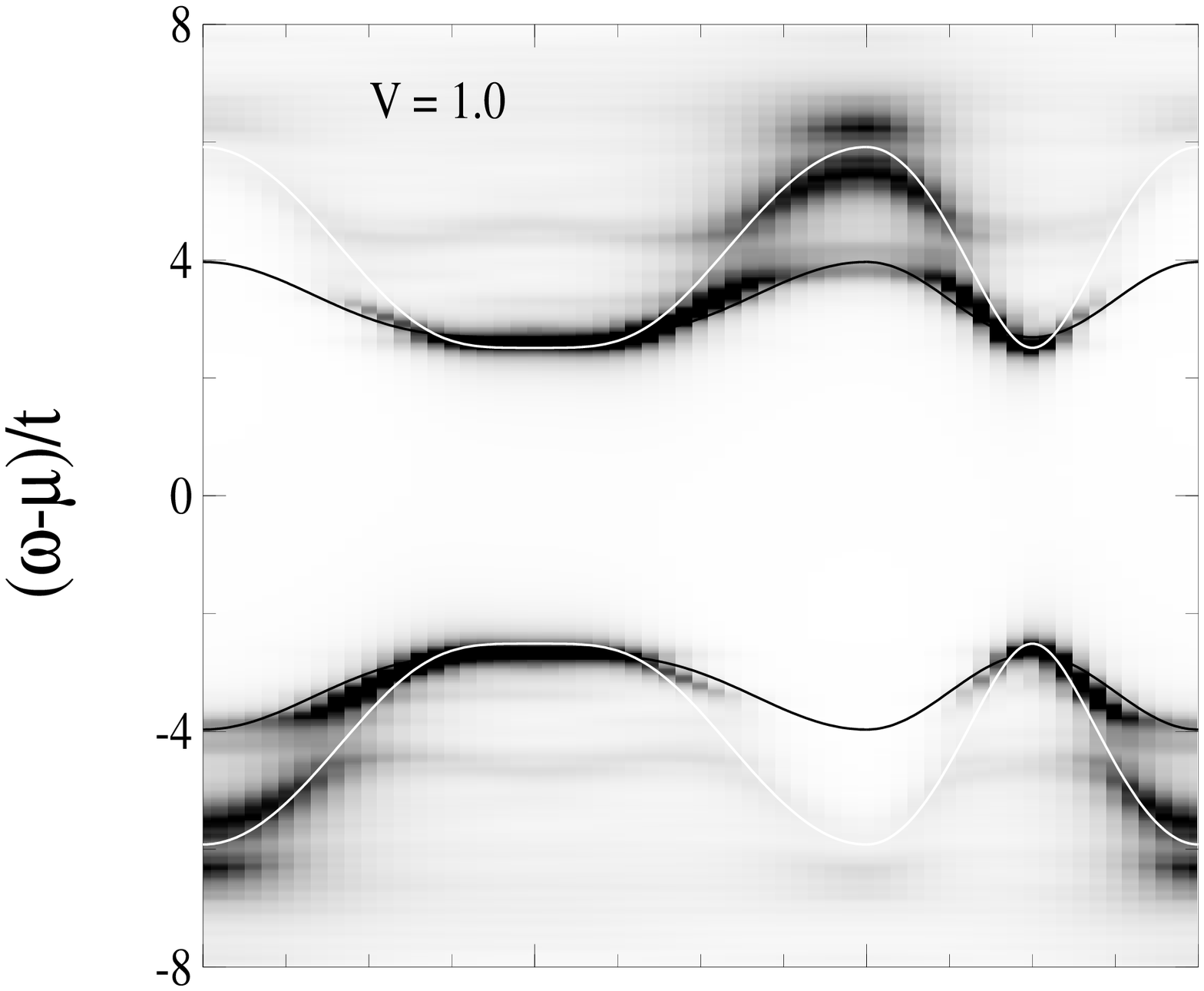}
  \includegraphics[width=0.45\textwidth,height=0.27\textheight]{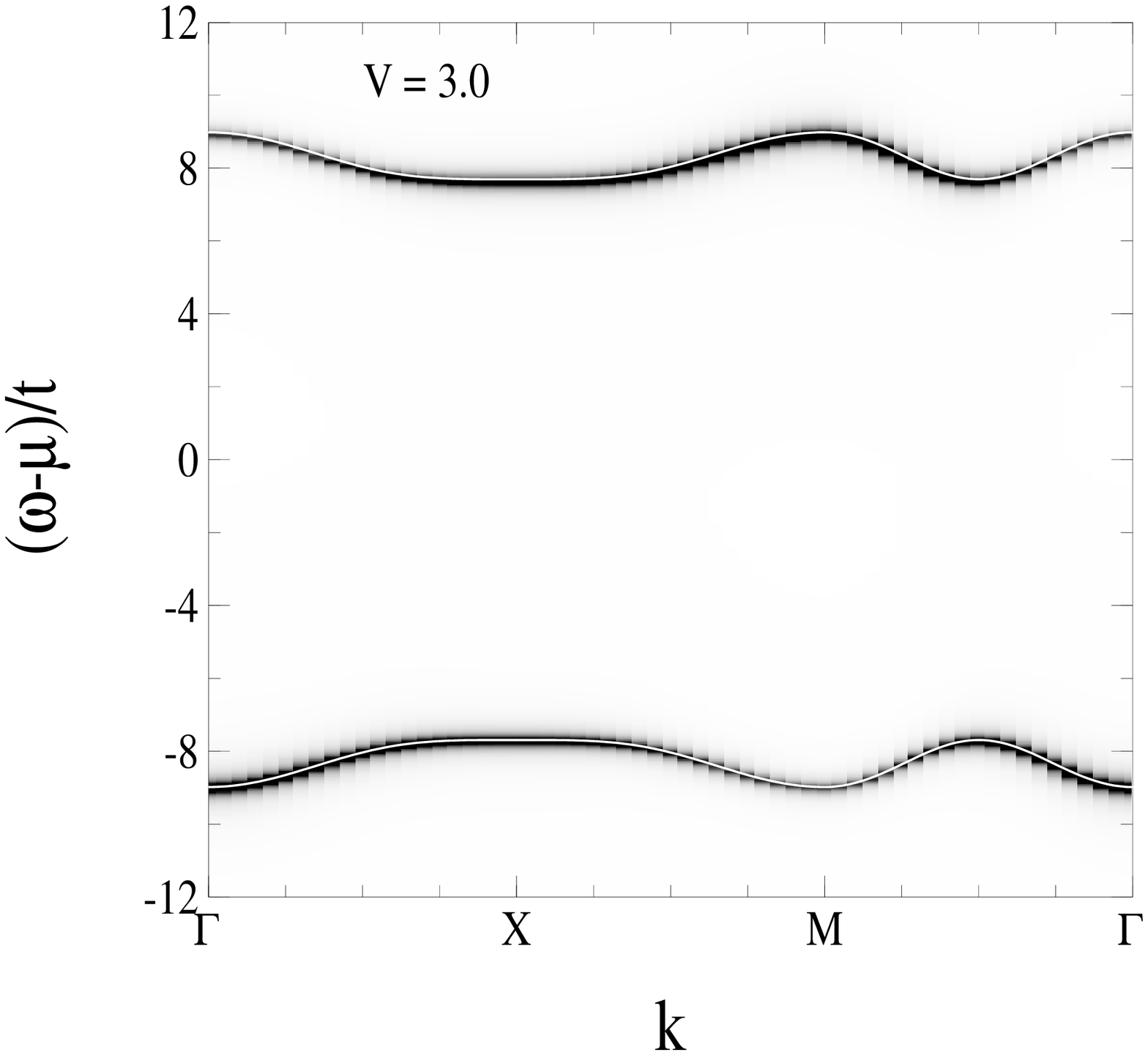}
  \caption{\label{fig:akw_2du8}%
    Density plot of the spectral function $A(\mathbf k,\omega)$ of the two-dimensional EHM at
    $U=8$ calculated on a $N_c=48$ super cluster with broadening
    $\sigma=0.1$. Darker regions represent larger spectral weight. Top: $V=1.0$. Bottom: 
    $V=3.0$. White lines are fits to Hartree-Fock dispersions. For the
    meaning of the black lines at $V=1.0$ see text.}
\end{figure}

\begin{figure}
  \centering
  \includegraphics[width=0.45\textwidth,height=0.24\textheight]{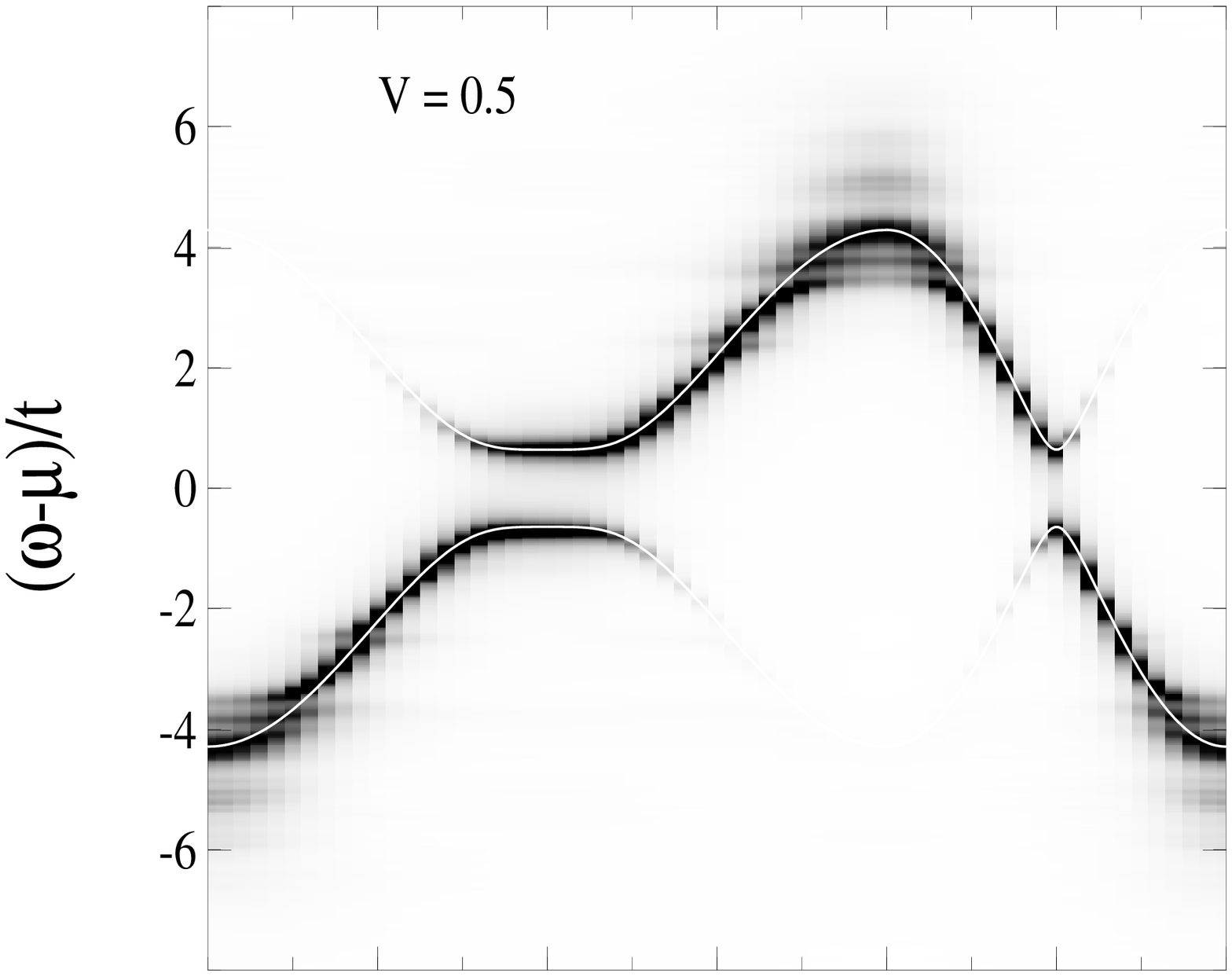}
  \includegraphics[width=0.45\textwidth,height=0.27\textheight]{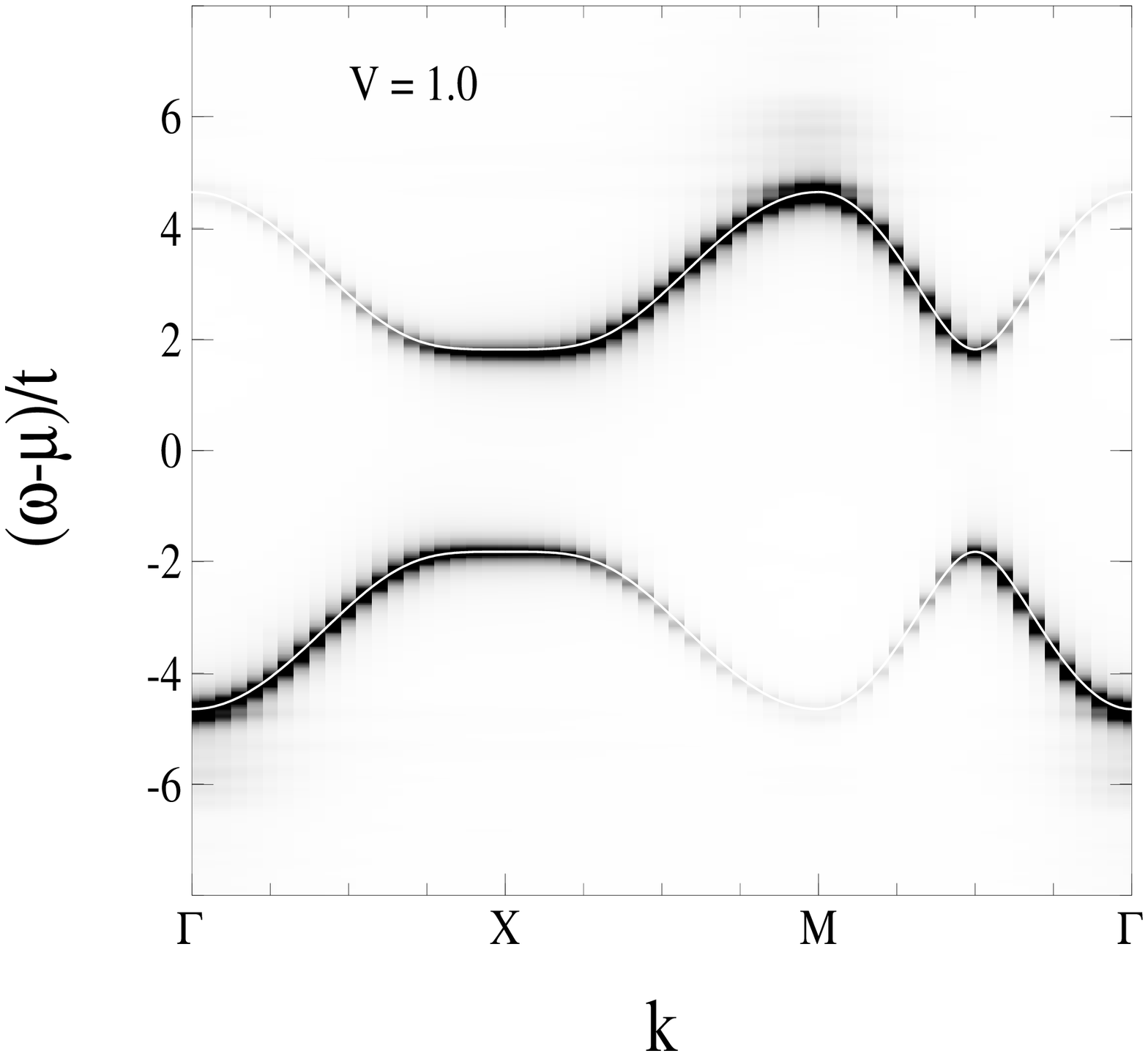}
  \caption{\label{fig:akw_2du3}%
    Same as \fig{fig:akw_2du8}, but at $U=3$. Top: $V=0.5$. Bottom: $V=1.0$.
  }
\end{figure}

The spectral function at $U=8$ in the SDW phase ($V=1.0$) and in the CDW
phase ($V=3.0$) is shown in \fig{fig:akw_2du8}. We found that the spectral
function at $V=1.0$ is very similar to the spectral function of the
Hubbard model ($V=0$).\cite{Dahnken03} One can see that the spectrum
mainly consists of four features, two high-energy Hubbard bands
and two low-energy quasi-particle bands, separated by a gap in the
spectrum. The dispersion of these low-energy excitations in 
the SDW phase differs significantly from the Hartree-Fock shape shown as white
lines in the upper panel of \fig{fig:akw_2du8}, which does not account for the
splitting into coherent low-energy bands and high-energy Hubbard bands. The
fit parameters were $t_{\rm fit}=1.34$ and $\Delta_{\rm fit}=2.51$.
The width of the coherent bands $|\omega(X)-\omega(\Gamma)|\approx 1.25$ is
rather set by the magnetic exchange $J$, consistent with QMC
calculations at $V=0$.\cite{Preuss95,Groeber00} The black lines are fits to
$E_{\mathbf k}=\pm\left[-\Delta+J/2(\cos k_x +\cos k_y)^2\right]$ which
accounts better for the dispersion of the low-energy bands than the Hartree-Fock
dispersion.\cite{Preuss95} The fit parameters 
were $\Delta_{\rm fit}=2.69$ and $J_{\rm fit}=-0.63$, which is in good agreement with the
second-order perturbation theory result $J=-4t^2/(U-V)=-0.57$. In the CDW
phase the white lines correspond to Hartree-Fock dispersions with fit parameters
$\Delta_{\rm fit}=7.69$ and $t_{\rm fit}=1.16$, and different from the SDW phase
they agree well with the excitations of $A({\mathbf k},\omega)$.  

\fig{fig:akw_2du3} displays the spectral function at $U=3$ at interactions
$V=0.5$ and $V=1.0$, respectively. The white lines again correspond
to Hartree-Fock dispersions with fit parameters
$t_{\rm fit}=1.06$, $\Delta_{\rm fit}=0.64$ at $V=0.5$, and $t_{\rm
  fit}=1.07$, $\Delta_{\rm fit}=1.82$ at $V=1.0$, respectively. As in the case $U=8$
the dispersion of the coherent low-energy bands in the SDW phase differs from
the Hartree-Fock prediction, but in this case 
the deviation is much smaller. We did not find an accurate functional form in
order to fit the low-energy excitations, but nevertheless we can extract the
value of $J$ from the band with of the coherent bands yielding
$J=-(1/2)|\omega(X)-\omega(\Gamma)|\approx -1.57$. This value is again in good
agreement with the perturbation theory result $J=-4t^2/(U-V)=-1.6$.

We would like to mention that our results at $U=3$ in the SDW phase are
qualitatively different from QMC results at $U=3$, $V=0$, and inverse
temperature $\beta=3t$,\cite{Groeber98} where the spectral function shows
metallic behavior with no gap around the Fermi energy. This difference may be
due to temperature effects or due to poor resolution of the Maximum-Entropy
inversion of QMC correlation functions. 

At both $U=8$ and $U=3$, one can easily see that agreement of the Hartree-Fock
dispersions with the low-energy excitations of $A({\mathbf
  k},\omega)$ is better in the CDW phase than in the SDW phase. In addition the
gap $\Delta_{\rm HF}$ calculated within the Hartree-Fock approximation is
much closer to the fitted gap $\Delta_{\rm fit}$ in the CDW phase
(e.g. $\Delta_{\rm HF}=7.76$, 
$\Delta_{\rm fit}=7.69$ at $U=8$, $V=3$) than in the SDW phase
(e.g. $\Delta_{\rm HF}=3.57$, $\Delta_{\rm fit}=2.51$ at $U=8$,
$V=0$). Therefore we conclude that in the CDW phase charge fluctuations play
only a minor role compared to the SDW phase, similar to the one-dimensional
system.

\section{Conclusions}\label{sec:concl}

In this paper we have presented a generalization of the variational cluster
perturbation theory to extended Hubbard models at half filling. The method is
based on the self-energy-functional approach (SFA) which uses dynamical information
of an exactly solvable system (reference system $H^\prime$) in order to 
approximate the physics in the thermodynamic limit. For the application 
of this method, a mean-field decoupling of the inter-cluster part of the 
nearest-neighbor Coulomb interaction is performed first. After this step, 
one is left with a Hamiltonian which couples the different clusters 
via the hopping only and which can be treated by the known (variational) 
CPT procedure.
The mean-field decoupling yields effective onsite potentials on the cluster
boundaries as external parameters of the Hamiltonian. These
parameters are determined either self-consistently on an isolated cluster 
(sufficient for the study of first order phase transitions) or by determination 
of the minimum of the SFA grand potential.

In order to test the accuracy of our approach we applied the method to the
extended Hubbard model in one dimension, because results from other methods
like QMC and DMRG are available for comparison. At $U=8$ the
results for the critical interaction $V_c$, the ground-state
energy, kinetic energy, and charge order parameter showed excellent
quantitative agreement with previous QMC studies. At $U=3$ our method
predicted a second-order phase transition with transition point $V_c=
1.665(5)$ again in good agreement with previous studies. The ground state
energy, kinetic energy, and order parameter do not seem to have been
calculated before.

In addition we calculated
the spectral function for several values of the interaction $V$, which has not
been done previously. At both $U=8$
and $U=3$, we found evidence for spin-charge separation in the SDW phase, but
not in the CDW phase. By fitting the bands by Hartree-Fock dispersions we found
that the hopping parameter is strongly renormalized. The
agreement between the fitted value of the gap and the value within the
Hartree-Fock approximation was much better in the CDW phase than in the SDW phase
giving rise to the conclusion that charge fluctuations play a minor role in
the CDW phase.

Whereas the application of sophisticated methods like DMRG or fermionic loop-update QMC
to more than one dimension is difficult, this extension is straightforward
within the present approach. We were thus able to perform the first
non-perturbative study of the
two-dimensional extended Hubbard model on a square lattice at half filling
and zero temperature beyond Hartree-Fock. We found first order transitions
at both $U=8$ and 
$U=3$ with transition points $V_c= 2.023(1)$ and $V_c= 0.770(3)$ for an
$N_c=48$ super cluster,
respectively. The spectral function in the SDW phase shows coherent
low-energy quasi-particle 
excitations with band width set by the magnetic exchange constant $J$, and
an incoherent background, consistent with previous QMC studies for the
Hubbard model at $V=0$. The Hartree-Fock prediction differs significantly
from the low-energy feature and does not describe the splitting into 
coherent quasi-particle bands and incoherent background. In the CDW phase the
Hartree-Fock dispersions account much better for 
the excitations, and no additional low-energy features caused by a
magnetic origin could be found. Similar to one dimension the agreement between
the Hartree-Fock approximation and the low-energy excitations obtained by the
present method is much better
in the CDW phase, confirming that charge fluctuations are less important in
the charge-ordered phase than in the SDW phase.

In this paper we applied our method to half-filled systems only, but one can
study ordering phenomena at other fillings, too, as long as the
possible order patterns are commensurate with the shape of clusters used as
reference system. With some effort it is also possible to study phases with
long wave-length charge density waves by coupling several clusters to a super
cluster and applying appropriate continuity conditions between the individual
clusters 
within the super cluster. In addition the application to systems with lattice
geometry different from the two-dimensional square lattice, e.g. ladder
materials, is an interesting subject for further studies. Work in this
direction is in progress.  

\begin{acknowledgements}

This work has been supported by the Austrian Science Fund (FWF), projects
P15834 and P15520.  
M. A. is supported by a doctoral scholarship program of the Austrian
Academy of Sciences. We gratefully acknowledge useful and stimulating
discussions with M. Hohenadler, W. Koller, E. Arrigoni, and C. Dahnken.

\end{acknowledgements}

\appendix

\section{Mean-field solution and free energy}\label{app:mf}

In this section we show that a mean-field solution obtained self
consistently is directly connected to a minimum in the free energy. For
simplicity let us assume that we have only two 
different mean-field parameters $\lambda_A=1-\delta$ and $\lambda_B=1+\delta$, see
also \sect{sec:decoupling}. We can write the mean-field decoupled
Hamiltonian \eq{eq:ehm_mf_decoupled} as
\begin{equation}
  H_{\rm MF}(\delta)=H_{\rm MF}^{(0)}+\sum_\R H_{\rm MF}^{(1)}(\R,\delta),
\end{equation}
where $H_{\rm MF}^{(0)}$ includes all terms independent of the mean-field
parameters. According to the third line in \eq{eq:ehm_mf_approx}, $H_{\rm
  MF}^{(1)}(\R,\delta)$ is given by
\begin{align}
  &H_{\rm MF}^{(1)}(\R,\delta)=V\sum_{[ij]}\left[ n_{\R i}\lambda_B + n_{\R
      j}\lambda_A - \lambda_A \lambda_B \right]\nonumber\\
  &=V\sum_{[ij]}\left[ n_{\R i}(1+\delta)+ n_{\R j}(1-\delta) - (1-\delta^2)\right]
\end{align}
where we assumed without loss of generality that the bonds $[ij]$ connect
sites $i$ on sublattice $A$ with sites $j$ on sublattice $B$.
The free energy of the system is given by
\begin{align}
  F&=-\frac{1}{\beta}\ln Z=-\frac{1}{\beta}\ln\tr e^{-\beta H_{\rm MF}(\delta)}\nonumber\\
  &=-\frac{1}{\beta}\ln\tr\exp\left[-\beta H_{\rm MF}^{(0)}-\beta\sum_\R H_{\rm MF}^{(1)}(\R,\delta)\right].
\end{align}
Taking the derivative with respect to $\delta$ yields
\begin{equation}
  \frac{\partial F}{\partial\delta}=V\sum_\R\left\langle\sum_{[ij]}\left[ n_{\R i} -
      n_{\R j} + 2\delta\right] \right\rangle.
\end{equation}
All clusters are equivalent, therefore we suppress the index $\R$ in the
following. Setting this derivative to zero we get the self-consistency
condition
\begin{equation}\label{eq:app:self_consist}
  \sum_{[ij]}\left[\langle n_i\rangle -\langle n_j\rangle +2\delta\right] =0.
\end{equation}
For one dimension, \eq{eq:app:self_consist} is given by
\begin{equation}
  \langle n_N\rangle -\langle n_1\rangle = 2\delta,
\end{equation}
because in this case we have only one decoupled bond $[1N]$ with site 1(N)
belonging to sublattice $A(B)$, respectively.
To conclude, one can state that if self consistency, \eq{eq:app:self_consist}, is
fulfilled, then the free energy has an extremum with respect to the mean-field
parameter $\delta$. By thermodynamic stability arguments this extremum always
has to be a minimum.

\end{document}